\documentclass[a4paper]{article}

\usepackage{cite}
\usepackage{graphicx}
\usepackage[ruled]{algorithm}
\usepackage{algpseudocode}
\usepackage{amsmath, amssymb, theorem}
\usepackage{subfigure}
\usepackage{multirow}
\usepackage{booktabs}
\usepackage{times}
\usepackage{longtable}

\usepackage{setspace}

\usepackage{caption}    

\usepackage{sectsty}
\sectionfont{\nohang\raggedright\normalsize\bfseries\MakeUppercase}
\subsectionfont{\raggedright\mdseries\normalsize\MakeUppercase}
\subsubsectionfont{\raggedright\mdseries\normalsize}

\makeatletter
\def\@seccntformat#1{\csname the#1\endcsname.\quad}
\makeatother

\theoremheaderfont{\scshape} \theoremstyle{plain}

\newcommand{\R}{\mathbb{R}}

\renewcommand{\eqref}[1]{(\ref{#1})}

\begin{document}
\doublespacing

\title{\textbf{\normalsize MACS: AN AGENT-BASED MEMETIC MULTIOBJECTIVE OPTIMIZATION ALGORITHM APPLIED TO SPACE TRAJECTORY DESIGN}}
\author{
    \normalsize Massimiliano Vasile -- Federico Zuiani \\
    \normalsize Space Advanced Research Team, University of Strathclyde \\
    \normalsize  James Weir Building
75 Montrose Street G1 1XJ, Glasgow, United Kingdom\\
    \normalsize massimiliano.vasile@strath.ac.uk, f.zuiani@aero.gla.ac.uk}
\date{}

\maketitle

\section*{abstract}
This paper presents an algorithm for multiobjective optimization
that blends together a number of heuristics. A population of agents
combines heuristics that aim at exploring the search space both
globally and in a neighborhood of each agent. These heuristics are
complemented with a combination of a local and global archive. The
novel agent-based algorithm is tested at first on a set of standard
problems and then on three specific problems in space trajectory
design. Its performance is compared against a number of
state-of-the-art multiobjective optimisation algorithms that use the
Pareto dominance as selection criterion: NSGA-II, PAES, MOPSO, MTS.
The results demonstrate that the agent-based search can identify
parts of the Pareto set that the other algorithms were not able to
capture. Furthermore, convergence is statistically better although the
variance of the results is in some cases higher.

\section*{keywords}

Multiobjective optimisation; trajectory optimisation; memetic algorithms; multiagent systems.

\section{Introduction}\label{sec:1}

The design of a space mission steps through different phases of
increasing complexity. In the first phase, a trade-off analysis of
several options is required. The trade-off analysis compares and
contrasts design solutions according to different criteria and aims
at selecting one or two options that satisfy mission requirements.
In mathematical terms, the problem can be formulated as a
multiobjective optimization problem.

As part of the trade-off analysis, multiple transfer trajectories to
the destination need to be designed. Each transfer should be optimal
with respect to a number of criteria. The solution of the associated
multiobjective optimization problem, has been addressed, by many
authors, with evolutionary techniques. Coverstone et al.
\cite{coverstone:00} proposed the use of multiobjective genetic
algorithms for the optimal design of low-thrust trajectories.
Dachwald et al. proposed the combination of a neurocontroller and of
a multiobjective evolutionary algorithm for the design of low-thrust
trajectories\cite{dachwald:04}. In 2005 a study by Lee et al.
\cite{lee:05} proposed the use of a Lyapunov controller with a
multiobjective evolutionary algorithm for the design of low-thrust
spirals. More recently, Sch\"{u}tze et al. proposed some innovative
techniques to solve multiobjective optimization problems for
multi-gravity low-thrust trajectories. Two of the interesting
aspects of the work of Sch\"{u}tze et al. are the archiving of
$\epsilon$- and $\Delta$-approximated solutions, to the known best
Pareto front \cite{schuetze:08c}, and the deterministic pre-pruning
of the search space\cite{schuetze:09}. In 2009, Delnitz et al.\
\cite{delnitz:09} proposed the use of multiobjective subdivision
techniques for the design of low-thrust transfers to the halo orbits
around the $L_2$ libration point in the Earth-Moon system. Minisci
et al. presented an interesting comparison between an EDA-based
algorithm, called MOPED, and NSGA-II on some constrained and
unconstrained multi-impulse orbital transfer problems
\cite{minisci:09}.

In this paper, a hybrid population-based approach that blends a
number of heuristics is proposed. In particular, the search for
Pareto optimal solutions is carried out globally by a population of
agents implementing classical social heuristics and more locally by
a subpopulation implementing a number of individualistic actions.
The reconstruction of the set of Pareto optimal solutions is handled
through two archives: a local and a global one.


The individualistic actions presented in this paper are devised to
allow each agent to independently converge to the Pareto optimal
set. Thus creating its own partial representation of the Pareto
front. Therefore, they can be regarded as memetic mechanisms
associated to a single individual. It will be shown that
individualistic actions significantly improve the performance of the
algorithm.

The algorithm proposed in this paper is an extension of the
Multi-Agent Collaborative Search (MACS), initially proposed in
\cite{vasile_cec:07, macs_jogo:08}, to the solution of
multiobjective optimisation problems. Such an extension required the
modification of the selection criterion, for both global and local
moves, to handle Pareto dominance and the inclusion of new
heuristics to allow the agents to move toward and along the Pareto
front. As part of these new heuristics, this papers introduces a
dual archiving mechanism for the management of locally and globally
Pareto optimal solutions and an attraction mechanisms that improves
the convergence of the population.

The new algorithm is here applied to a set of known standard test
cases and to three space mission design problems. The space mission
design cases in this paper consider spacecraft equipped with a
chemical engine and performing a multi-impulse transfer. Although
these cases are different from some of the above-mentioned examples,
that consider a low-thrust propulsion system, nonetheless the size
and complexity of the search space is comparable. Furthermore,
it provides a first test benchmark for multi-impulsive problems that
have been extensively studied in the single objective case but for
which only few comparative studies exist in the multiobjective case
\cite{minisci:09}.

The paper is organised as follows: section two contains the general
formulation of the problem, the third section starts with a general
introduction to the multi-agent collaborative search algorithm and
heuristics before going into some of the implementation details.
Section four contains a set of comparative tests that demonstrates
the effectiveness of the heuristics implemented in MACS. The section
briefly introduces the algorithms against which MACS is compared and
the two test benchmarks that are used in the numerical experiments.
It then defines the performance metrics and ends with the results of
the comparison.

\section{Problem Formulation}
\label{sec:Problem} A general problem in multiobjective optimization
is to find the feasible set of solutions that satisfies the
following problem:
\begin{equation}\label{eq:gen_MOO_prob}
    \min_{\mathbf{x}\in D} \mathbf{f}(\mathbf{x)}
\end{equation}
where $D$ is a hyperrectangle defined as $D = \big\lbrace x_j \mid
x_j \in [b_j^l \; b_j^u] \subseteq \R, \; j=1,...,n \big\rbrace$ and
$\mathbf{f}$ is the vector function:
\begin{equation}\label{eq:gen_F}
    \mathbf{f}:D \rightarrow \R^m,
    \;\;\;\;\
    \mathbf{f}(\mathbf{x)}=[f_1(\mathbf{x}),f_2(\mathbf{x}),...,f_m(\mathbf{x})]^T
\end{equation}

The optimality of a particular solution is defined through the
concept of dominance: with reference to problem
(\ref{eq:gen_MOO_prob}), a vector $\mathbf{y}\in D$ is dominated by
a vector $\mathbf{x}\in D$ if $f_j(\mathbf{x})<f_j(\mathbf{y})$ for
all $j=1,...,m$. The relation $\mathbf{x} \prec \mathbf{y}$ states
that $\mathbf{x}$ dominates $\mathbf{y}$.

Starting from the concept of dominance, it is possible to associate,
to each solution in a set, the scalar dominance index:
\begin{equation}\label{eq:dominance_index}
I_d(\textbf{x}_j)=|\{i\mid i\wedge j\in N_p \wedge
\textbf{x}_i\prec \textbf{x}_j\}|
\end{equation}
where the symbol $|.|$ is used to denote the cardinality of a set
and $N_p$ is the set of the indices of all the solutions. All
non-dominated and feasible solutions form the set:
\begin{equation}\label{eq:X_set}
X=\{\textbf{x}\in D \mid I_d(\textbf{x})=0\}
\end{equation}

Therefore, the solution of problem (\ref{eq:gen_MOO_prob})
translates into finding the elements of $X$. If $X$
is made of a collection of compact sets of finite measure in $\R^n$,
then once an element of $X$ is identified it makes sense to explore
its neighborhood to look for other elements of $X$. On the other
hand, the set of non dominated solutions can be disconnected and its
elements can form islands in $D$. Hence, restarting the search
process in unexplored regions of $D$ can increase the collection of
elements of $X$.

The set $X$ is the Pareto set and the corresponding image in
criteria space is the Pareto front. It is clear that in $D$ there
can be more than one $X_l$ containing solutions that are locally
non-dominated, or locally Pareto optimal. The interest is, however,
to find the set $X_g$ that contains globally non-dominated, or
globally Pareto optimal, solutions.

\section{Multiagent Collaborative Search}
\label{sec:macs} The key motivation behind the development of
multi-agent collaborative search was to combine local and global
search in a coordinated way such that local convergence is improved
while retaining global exploration \cite{macs_jogo:08}. This
combination of local and global search is achieved by endowing a set
of agents with a repertoire of actions producing either the sampling
of the whole search space or the exploration of a neighborhood of
each agent. More precisely, in the following, global exploration
moves will be called collaborative actions while local moves will be
called individualistic actions. Note that not all the
individualistic actions, described in this paper, aim at exploring a
neighborhood of each agent, though. The algorithm presented in this
paper is a modification of MACS to tackle multiobjective
optimization problems. In this section, the key heuristics
underneath MACS will be described together with their modification
to handle problem (\ref{eq:gen_MOO_prob}) and reconstruct $X$.

\subsection{General Algorithm Description}
A population $P_0$ of $n_{pop}$ virtual agents, one for each
solution vector $\textbf{x}_i$, with $i=1,...,n_{pop}$, is deployed
in $D$. The population evolves through a number of generations. At
every generation $k$, the dominance index (\ref{eq:dominance_index}) of each agent
$\textbf{x}_{i,k}$ in the population $P_k$ is evaluated. The agents with
dominance index $I_d=0$ form a set $X_k$ of non-dominated solutions.
Hence, problem (\ref{eq:gen_MOO_prob}) translates into finding a series of sets $X_k$ such that $X_k
\rightarrow X_g$ for $k\rightarrow k_{max}$ with $k_{max}$ possibly
a finite number.

The position of each agent in $D$ is updated through a number of
heuristics. Some are called \emph{collaborative actions} because are
derived from the interaction of at least two agents and involve the
entire population at one time. The general collaborative heuristic
can be expressed in the following form:
\begin{equation}\label{eq:col_heu}
    \mathbf{x}_{k}=\mathbf{x}_{k}+S(\mathbf{x}_{k}+\mathbf{u}_k)\mathbf{u}_k
\end{equation}
where $\mathbf{u}_k$ depends on the other agents in the population
and $S$ is a selection function which yields 0 if the candidate
point $\mathbf{x}_{k}+\mathbf{u}_k$ is not selected or 1 if it is
selected (see Section \ref{sec:collaboration}). In this
implementation a candidate point is selected if its dominance index
is better or equal than the one of $\mathbf{x}_{k}$. A first restart
mechanism is then implemented to avoid crowding. This restart
mechanism is taken from \cite{macs_jogo:08} and prevents the agents
from overlapping or getting too close. It is governed by the crowding
factor $w_c$ that defines the minimum acceptable normalized distance
between agents. Note that, this heuristic increases the uniform
sampling rate of $D$, when activated, thus favoring exploration. On
the other hand, by setting $w_c$ small the agents are more directed
towards local convergence.

After all the collaborative and restart actions have been
implemented, the resulting updated population $P_{k}$ is ranked
according to $I_d$ and split in two subpopulations: $P_{k}^u$ and
$P_{k}^l$. The agents in each subpopulation implement sets of, so
called, \emph{individualistic actions} to collect samples of the
surrounding space and to modify their current location. In
particular, the last $n_{pop}-f_en_{pop}$ agents belong to $P_{k}^u$
and implement heuristics that can be expressed in a form similar to
Eq. (\ref{eq:col_heu}) but with $\mathbf{u}_k$ that depends only on
$\mathbf{x}_{k}$.

The remaining $f_en_{pop}$ agents belong to $P_{k}^l$ and implement a mix of
actions that aim at either improving their location or exploring the neighborhood
$N_{\rho}(\mathbf{x}_{i,k})$, with $i=1,...,f_en_{pop}$.
$N_{\rho}(\mathbf{x}_{i,k})$ is a hyperectangle centered in
$\mathbf{x}_{i,k}$. The intersection
$N_{\rho}(\mathbf{x}_{i,k})\cap D$ represents the local
region around agent $\textbf{x}_{i,k}$ that one wants to explore. The {\em size} of
$N_{\rho}(\mathbf{x}_{i,k})$ is specified by the value
$\rho(\textbf{x}_{i,k})$: the $i^\text{th}$ edge of $N_{\rho}$ has length
$2\rho(\textbf{x}_{i,k}) \max\{b_j^u-x_{i,k}[j],
x_{i,k}[j]-b^l_j\}$.

The agents in $P_{k}^l$ generate a number of perturbed solutions
$\mathbf{y}_s$ for each $\textbf{x}_{i,k}$, with $s=1,...,s_{max}$.
These solutions are collected in a local archive $A_l$ and a
dominance index is computed for all the elements in $A_l$. If at
least one element $\mathbf{y}_s\in A_l$ has $I_d=0$ then
$\mathbf{x}_{i,k}\leftarrow \mathbf{y}_s$. If multiple elements of
$A_l$ have $I_d=0$, then the one with the largest variation, with
respect to $\mathbf{x}_{i,k}$, in criteria space is taken. Figure
\ref{fig:local} shows three agents (circles) with a set of locally
generated samples (stars) in their respective neighborhoods (dashed
square). The arrows indicate the direction of motion of each agent.
The figures shows also the local archive for the first agent
$A_{l_1}$ and the target global archive $X_g$.

The $\rho$ value associated to an agent is updated at each iteration
according to the rule devised in \cite{macs_jogo:08}. Furthermore, a
value $s(\textbf{x}_{i,k})$ is associated to each agent
$\textbf{x}_{i,k}$ to specify the number of samples allocated to the
exploration of $N_{\rho}(\mathbf{x}_{i,k})$. This value is updated
at each iteration according to the rule devised in
\cite{macs_jogo:08}.

The adaptation of $\rho$ is introduced to allow the agents to
self-adjust the neighborhood removing the need to set a priori the
appropriate size of $N_{\rho}(\mathbf{x}_{i,k})$. The consequence of
this adaptation is an intensification of the local search by some
agents while the others are still exploring. In this respect, MACS
works opposite to Variable Neighborhood Search heuristics, where the
neighborhood is adapted to improve global exploration, and
differently than Basin Hopping heuristics in which the neighborhood
is fixed. Similarly, the adaptation of $s(\textbf{x}_{i,k})$ avoids
setting a priori an arbitrary number of individualistic moves and
has the effect of avoiding an excessive sampling of
$N_{\rho}(\mathbf{x}_{i,k})$ when $\rho$ is small. The value of
$s(\textbf{x}_{i,k})$ is initialized to the maximum number of
allowable individualistic moves $s_{max}$. The value of $s_{max}$ is
here set equal to the number of dimensions $n$. This choice is
motivated by the fact that a gradient-based method would evaluate
the function a minimum of $n$ times to compute an approximation of
the gradient with finite differences. Note that the set of
individualistic actions allows the agents to independently move
towards and within the set $X$, although no specific mechanism is
defined as in \cite{schuetze_hcs:08}. On the other hand the
mechanisms proposed in \cite{schuetze_hcs:08} could further improve
the local search and will be the subject of a future investigation.

 All the elements of
$X_k$ found during one generation are stored in a global archive
$A_g$. The elements in $A_g$ are a selection of the elements
collected in all the local archives. Figure \ref{fig:global}
illustrates three agents performing two social actions that yield
two samples (black dots). The two samples together with the
non-dominated solutions coming from the local archive form the
global archive. The archive $A_g$ is used to implement an attraction
mechanism that improves the convergence of the worst agents (see
Sections \ref{sec:att}). 
During the global archiving process a second restart mechanism that
reinitializes a portion of the population (\emph{bubble restart}) is
implemented. Even this second restart mechanism is taken from
\cite{macs_jogo:08} and avoids that, if $\rho$ collapses to zero, the
agent keeps on sampling a null neighborhood.

Note that, within the MACS framework, other strategies can be
assigned to the agents to evaluate their moves in the case of
multiple objective functions, for example a decomposition
technique\cite{MOEAD07}. However, in this paper we develop the
algorithm based only on the use of the dominance index.

\begin{figure}
\begin{center}
  \caption{Illustration of the a) local moves and archive and b) global moves and archive.}\label{fig:local_global}
  \end{center}
\end{figure}

\begin{algorithm}[!ht]
\caption {Main MACS algorithm} \label{alg:MACS}
\begin{algorithmic}[1]

\State Initialize a population $P_0$ of $n_{pop}$ agents in $D$, $k=0$, number of function evaluations $n_{eval}=0$, maximum number of function evaluations $N_e$, crowding factor $w_c$

\ForAll{$i=1,...,n_{pop}$}

\State
$\mathbf{x}_{i,k}=\mathbf{x}_{i,k}+S(\mathbf{x}_{i,k}+\mathbf{u}_{i,k})\mathbf{u}_{i,k}$
\EndFor
\State Rank solutions in $P_{k}$ according to $I_d$

\State Re-initialize crowded agents according to the
single agent restart mechanism

\ForAll{$i=f_en_{pop},...,n_{pop}$}

\State Generate $n_p$ mutated copies of $\mathbf{x}_{i,k}$.
\State Evaluate the dominance of each mutated copy $\mathbf{y}_p$
against $\mathbf{x}_{i,k}$, with $p=1,...,n_p$

\If{$\exists p| \mathbf{y}_p\prec\mathbf{x}_{i,k}$}

\State $\bar{p} = \arg \max_p \| \mathbf{y}_p-\mathbf{x}_{i,k} \|$

\State $\mathbf{x}_{i,k} \leftarrow \mathbf{y}_{\bar{p}}$

\EndIf

 \EndFor

\ForAll{$i=1,...,f_en_{pop}$}

\State Generate $s<s_{max}$ individual actions $\mathbf{u}_s$ such
that $\mathbf{y}_s=\mathbf{x}_{i,k}+\mathbf{u}_s$

\If{$\exists s| \mathbf{y}_s\prec \mathbf{x}_{i,k}$}

\State $\bar{s} = \arg \max_s \| \mathbf{y}_s-\mathbf{x}_{i,k} \|$

\State $\mathbf{x}_{i,k}\leftarrow \mathbf{y}_{\bar{s}}$ \EndIf

\State Store candidate elements $\mathbf{y}_s$ in the local archive
$A_l$

\State Update $\rho(\mathbf{x}_{i,k})$ and $s(\mathbf{x}_{i,k})$

\EndFor

\State Form $P_k=P_k^l\bigcup P_k^u$ and $\hat{A}_g=A_g\bigcup A_l\bigcup P_k$
\State Compute $I_d$ of all the elements in $\hat{A}_g$

\State $A_g=\{\mathbf{x}|\mathbf{x}\in \hat{A}_g \wedge
I_d(\mathbf{x})=0\wedge \|\mathbf{x}-\mathbf{x}_{A_g}\|>w_c\}$

\State Re-initialize crowded
agents in $P_{k}$ according to the second restart mechanism

\State Compute attraction component to $A_g$ for all
$\mathbf{x}_{i,k}\in P_k\setminus X_k$ \State $k=k+1$
 \State{\textbf{Termination}} Unless $n_{eval}>N_{e}$, GoTo
 Step~2

\end{algorithmic}
\end{algorithm}

\subsection{Collaborative Actions} \label{sec:collaboration}
Collaborative actions define operations through which information is
exchanged between pairs of agents. Consider a pair of agents
$\textbf{x}_1$ and $\textbf{x}_2$, with $\textbf{x}_1\prec
\mathbf{x}_2$. One of the two agents is selected at random in the
worst half of the current population (from the point of view of the
property $I_d$), while the other is selected at random from the
whole population. Then, three different actions are performed. Two
of them are defined by adding to $\textbf{x}_1$ a step $\mathbf{u}_k$ defined as follows:
\begin{equation}
\mathbf{u}_k=\alpha
r^t(\textbf{x}_2-\textbf{x}_1),
\end{equation}
and corresponding to: {\em extrapolation} on the side of ${\bf x}_1$ ($\alpha=-1$, $t=1$), with the further constraint that the result must belong to the domain $D$ (i.e., if the step
$\mathbf{u}_k$ leads out of $D$, its size is reduced until we get back to $D$); {\em interpolation} ($\alpha=1$), where a random point between $\textbf{x}_1$ and $\textbf{x}_2$ is sampled. In the latter case, the shape parameter $t$ is defined as follows:
\begin{equation}
t=0.75\frac{s({\bf x}_1)-s({\bf x}_2)}{s_{\max}}+1.25
\end{equation}
The rationale behind this definition is that we are favoring moves
which are closer to the agent with a higher fitness value if the two
agents have the same $s$ value, while in the case where the agent
with highest fitness value has a $s$ value much lower than that of
the other agent, we try to move away from it because a small $s$
value indicates that improvements close to the agent are difficult
to detect. \newline The third operation is the {\em recombination}
operator, a {\em single-point crossover}, where, given the two
agents: we randomly select a component $j$; split the two agents
into two parts, one from component 1 to component $j$ and the other
from component $j+1$ to component $n$; and then we combine the two
parts of each of the agents in order to generate two new solutions.
The three operations give rise to four new samples, denoted by
$\textbf{y}_1$, $\textbf{y}_2$, $\textbf{y}_3$, $\textbf{y}_4$.
Then, $I_d$ is computed for the set ${\mathbf{x}_{2},\textbf{y}_1,
\textbf{y}_2, \textbf{y}_3, \textbf{y}_4}$. The element with $I_d=0$
becomes the new location of $\mathbf{x}_2$ in $D$.

\subsection{Individualistic Actions} \label{sec:beh}

Once the collaborative actions have been implemented,
each agent in $P_{k}^u$ is mutated a number of times: the lower the
ranking the higher the number of mutations. The mutation mechanisms is not different from the single objective case but the selection is modified to use the dominance index rather than the objective values.

A mutation is simply a random vector $\mathbf{u}_{k}$ such that $\mathbf{x}_{i,k}+\mathbf{u}_{k}\in D$.
All the mutated solution
vectors are then compared to $\mathbf{x}_{i,k}$, i.e. $I_d$ is
computed for the set made of the mutated solutions and
$\mathbf{x}_{i,k}$. If at least one element $\mathbf{y}_p$ of the
set has $I_d=0$ then $\mathbf{x}_{i,k}\leftarrow \mathbf{y}_p$. If
multiple elements of the set have $I_d=0$, then the one with the
largest variation, with respect to $\mathbf{x}_{i,k}$, in criteria
space is taken.

Each agent in $P_{k}^l$ performs at most $s_{max}$ of the following individualistic actions: \emph{inertia}, \emph{differential}, \emph{random with line search}. The overall procedure is summarized in Algorithm
\ref{alg:individual} and each action is described in detail in the following subsections.

\subsubsection{Inertia}
If agent $i$ has improved from generation $k-1$ to generation $k$,
then it follows the direction of the improvement (possibly until it
reaches the border of $D$), i.e., it performs the following step:
\begin{equation}\label{13}
\mathbf{y}_s =\textbf{x}_{i,k}+\bar{\lambda}\Delta_I
\end{equation}
where $ \bar{\lambda}=\min\{1,\max\{\lambda\ :\ \mathbf{y}_s \in
D\}\}$ and $\Delta_I=(\textbf{x}_{i,k}-\textbf{x}_{i,k-1})$.

\subsubsection{Differential}
This step is inspired by Differential Evolution\cite{kenneth:05}. It is defined as follows: let $\textbf{x}_{i_1,k}, \textbf{x}_{i_2,k},
\textbf{x}_{i_3,k}$ be three randomly selected agents; then
\begin{equation}
\label{eq:demove}
\mathbf{y}_{s}=\textbf{x}_{i,k}+\mathbf{e}\left[\textbf{x}_{i_1,k}+F(\textbf{x}_{i_3,k}-\textbf{x}_{i_2,k})\right]
\end{equation}
with $\mathbf{e}$ a vector containing a random number of 0 and 1
(the product has to be intended componentwise) with probability 0.8 and $F=0.8$ in this implementation. For every component
$y_s[j]$ of $\mathbf{y}_s$ that is outside the boundaries defining $D$ then $y_s[j]=r(b^u_j-b^l_j)+b^l_j$, with $r\in U(0,1)$.
Note that, although this action involves more than one agent, its outcome is only compared to the other outcomes coming from the actions performed by agent $\textbf{x}_{i,k}$ and therefore it is considered individualistic.

\subsubsection{Random with Line Search}
This move realizes a local exploration of the neighborhood $N(\mathbf{x}_{i,k})$. It generates a first random sample $\mathbf{y}_s\in
N(\mathbf{x}_{i,k})$. Then if $\mathbf{y}_s $ is not an improvement, it generates a second sample
$\mathbf{y}_{s+1}$ by extrapolating on the side of the better one
between $\mathbf{y}_s$ and $\mathbf{x}_{i,k}$:
\begin{equation}\label{15}
\mathbf{y}_{s+1}=\mathbf{x}_{i,k}+\bar{\lambda}\left[\alpha_2
r^t(\textbf{y}_s-\textbf{x}_{i,k})+\alpha_1(\textbf{y}_s-\textbf{x}_{i,k})\right]
\end{equation}
with $ \bar{\lambda}=\min\{1,\max\{\lambda\ :\ \mathbf{y}_{s+1} \in D\}\}$ and where $\alpha_1,\alpha_2\in \{-1,0,1\}$, $r\in U(0,1)$ and $t$ is a shaping parameter which controls the magnitude of the displacement. Here we use the parameter values $\alpha_1=0$, $\alpha_2=-1$, $t=1$, which corresponds to {\em extrapolation} on the side of $\textbf{x}_{i,k}$, and $\alpha_1=\alpha_2=1$, $t=1$, which corresponds to {\em extrapolation} on the side of $\textbf{y}_s$.

The outcome of the extrapolation is used to construct a second order one-dimensional model of $I_d$. The second order model is given by the quadratic function $f_l(\sigma)=a_1 \sigma^2+a_2\sigma+a_3$ where $\sigma$ is a coordinate along the $\mathbf{y}_{s+1}-\mathbf{y}_{s}$ direction. The coefficients $a_1$, $a_2$ and $a_3$ are computed so that $f_l$ interpolates the values $I_d(\mathbf{y}_s)$, $I_d(\mathbf{y}_{s+1})$ and $I_d(\mathbf{x}_{i,k})$. In particular, for $\sigma=0$ $f_l=I_d(\mathbf{y}_{s})$, for $\sigma=1$ $f_l=I_d(\mathbf{y}_{s+1})$ and for $\sigma=(\mathbf{x}_{i,k}-\mathbf{y}_s)/\|\mathbf{x}_{i,k}-\mathbf{y}_s\|$
$f_l=I_d(\mathbf{x}_{i,k})$. Then, a new sample $\mathbf{y}_{s+2}$ is taken at the minimum of the second-order model along the $\mathbf{y}_{s+1}-\mathbf{y}_{s}$ direction.

\begin{algorithm}[!ht]
\caption {Individual Actions in $P_k^l$} \label{alg:individual}
\begin{algorithmic}[1]
\State $s=1$, $stop=0$

\If{$\mathbf{x}_{i,k}\prec \mathbf{x}_{i,k-1}$}

$\mathbf{y}_s
=\textbf{x}_{i,k}+\bar{\lambda}(\textbf{x}_{i,k}-\textbf{x}_{i,k-1})$

with $ \bar{\lambda}=\min\{1,\max\{\lambda\ :\ \textbf{y}_{s}\in D
\}\}.$ \EndIf

\If{$ \mathbf{x}_{i,k} \prec \textbf{y}_{s}$}

 $s=s+1$

$
\textbf{y}_s=\mathbf{e}[\textbf{x}_{i,k}-(\textbf{x}_{i_1,k}+(\textbf{x}_{i_3,k}-\textbf{x}_{i_2,k}))]$

$\forall j| y_s(j) \notin D$,
$y_s(j)=r(b_u(j)-b_l(j))+b_l(j)$,

with $j=1,...,n$ and $r\in U(0,1)$

\Else $\quad stop=1$
 \EndIf

\If{$ \mathbf{x}_{i,k} \prec \mathbf{y}_s$}

$s=s+1$

Generate $\textbf{y}_s\in N_{\rho}(\mathbf{x}_{i,k})$.

Compute
$\mathbf{y}_{s+1}=\mathbf{x}_{i,k}+\bar{\lambda}r^t(\textbf{x}_{i,k}-\textbf{y}_s)$

with $ \bar{\lambda}=\min\{1,\max\{\lambda\ :\ \textbf{y}_s\in D
\}\}$

and $r\in U(0,1)$.

Compute
$\mathbf{y}_{s+2}=\bar{\lambda}\sigma_{min}(\textbf{y}_{s+1}-\textbf{y}_{s})/\|(\textbf{y}_{s+1}-\textbf{y}_{s})\|$,

with $\sigma_{min}= \arg\min_{\sigma} \{a_1 \sigma^2+a_2
\sigma+I_d(\mathbf{y}_{s})\}$,

and $ \bar{\lambda}=\min\{1,\max\{\lambda\ :\ \textbf{y}_{s+2}\in D
\}\}.$

$s=s+2$

 \Else {$\quad stop=1$}

 \EndIf

 \State{\textbf{Termination}} Unless $s>s_{max}$ or $stop=1$ , GoTo
 Step~4

\end{algorithmic}
\end{algorithm}

The position of
$\textbf{x}_{i,k}$ in $D$ is then updated with the $\mathbf{y}_s$ that has
$I_d=0$ and the longest vector difference in the criteria space with
respect to $\textbf{x}_{i,k}$. The displaced vectors
$\mathbf{y}_s$ generated by the agents in $P_{k}^l$ are not discarded but contribute to a local archive $A_l$, one for each agent, except for the one selected to update the location of $\textbf{x}_{i,k}$. In order to rank the $\mathbf{y}_s$, the following modified dominance index is used:
\begin{equation}\label{e18c}
\begin{array}{c}
\hat{I}_d(\textbf{x}_{i,k}) = \Big| \big\lbrace  j \mid
f_j(\textbf{y}_{s}) = f_j(\textbf{x}_{i,k}) \big\rbrace \Big| \kappa
+\\
 \Big| \big\lbrace j \mid f_j(\textbf{y}_{s}) >
f_j(\textbf{x}_{i,k})  \big\rbrace \Big|
\end{array}
\end{equation}
where $\kappa$ is equal to one if there is at least one component of
$\textbf{f}(\textbf{x}_{i,k})=[f_1,f_2,...,f_m]^T$ which is better
than the corresponding component of $\textbf{f}(\textbf{y}_{s})$,
and is equal to zero otherwise.

Now, if for the ${s}^{th}$ outcome, the dominance index in Eq.\ \eqref{e18c} is not zero but is lower than the number of components of the objective vector, then the agent $\textbf{x}_{i,k}$ is only partially dominating the $s^\text{th}$ outcome. Among all the partially dominated outcomes with the same dominance index we select the one that satisfies the condition:
\begin{equation}\label{e18e}
\min\limits_{s} \big \langle \big(
\textbf{f}(\textbf{x}_{i,k}))-\textbf{f}(\textbf{y}_{s}) \big),
\textbf{e} \big \rangle
\end{equation}
where $\textbf{e}$ is the unit vector of dimension $m$, $\textbf{e}=\frac{[1,1,1,\ldots,1]^T}{\sqrt{m}}$. All the non-dominated and selected partially dominated solutions form the local archive $A_l$.

\subsection{The local and global archives $A_l$ and $A_g$}
\label{sec:arc}

Since the outcomes of one agent could dominate other agents or the
outcomes of other agents, at the end of each generation, every $A_l$
and the whole population $P_k$ are added to the current global
archive $A_g$. The global $A_g$ contains $X_k$, the current best
estimate of $X_g$. The dominance index in Eq.(\ref{eq:dominance_index}) is then
computed for all the elements in $\hat{A}_g=A_g\bigcup_l A_l\bigcup
P_k$ and only the non-dominated ones with crowding distance
$\|\mathbf{x}_{i,k}-\mathbf{x}_{A_g}\|>w_c$ are preserved (where
$\mathbf{x}_{A_g}$ is an element of $A_g$).


\subsubsection{Attraction} \label{sec:att}
The archive $A_g$ is used to direct the movements of those agents
that are outside $X_k$. All agents, for which $I_d\neq 0$ at step
$k$, are assigned the position of the elements in $A_g$ and their
inertia component is recomputed as:
\begin{equation}\label{e19e}
\Delta_I=r(\textbf{x}_{A_g}-\textbf{x}_{i,k})
\end{equation}
More precisely, the elements in the archive are ranked according to
their reciprocal distance or crowding factor. Then, every agent for
which $I_d\neq 0$ picks the least crowded element $\textbf{x}_{A_g}$
not already picked by any other agent.

\subsection{Stopping rule}
\label{sec:stop} The search
is stopped when a prefixed number $N_e$ of
function evaluations is reached. At termination of the
algorithm the whole final population is inserted into the archive
$A_g$.

\section{Comparative Tests}

The proposed optimization approach was
 implemented in a software code, in Matlab, called MACS. In previous works
 \cite{vasile_cec:07,macs_jogo:08}, MACS was tested on single objective
 optimization problems related to space trajectory design, showing
 good performances. In this work, MACS was tested at first
 on a number of standard problems, found in literature, and then on three
typical space trajectory optimization problems.

This paper extends the results presented in Vasile and Zuiani
(2010)\cite{MACS:10} by adding a broader suite of algorithms for
multiobjective optimization to the comparison and a different
formulation of the performance metrics.

\subsection{Tested Algorithms}
MACS was compared against a number of state-of-the-art algorithms
for multiobjective optimization. For this analysis it was decided to
take the basic version of the algorithms that is available online.
Further developments of the basic algorithms have not been
considered in this comparison and will be included in future works.
Note that, all the algorithms selected for this comparative analysis
use Pareto dominance as selection criterion.

The tested algorithms are: NSGA-II \cite{Deb:00},
MOPSO\cite{MOPSO:01}, PAES\cite{PAES:99} and MTS\cite{MTS:07}. A
short description of each algorithm with their basic parameters
follows.

\subsubsection{NSGA-II}
The Non-Dominated Sorting Genetic Algorithm (NSGA-II) is a genetic algorithm which uses the concept of dominance class (or depth) to rank the population. A crowding factor is then used to rank the individuals within each dominance class. Optimization starts from a randomly generated initial population. The individuals in the population are sorted according to their level of Pareto dominance with respect to other individuals. To be more precise, a fitness value equal to 1 is assigned to the non-dominated individuals. Non-dominated individuals form the first layer (or class). Those individuals dominated only by members of the first layer form the second one and are assigned a fitness value of 2, and so on. In general, for dominated individuals, the fitness is given by the number of dominating layers plus 1.
A crowding factor is then assigned to each individual in a given class. The crowding factor is computed as the sum of the Euclidean distances, in criteria space, with respect to other individuals in the same class, divided by the interval spanned by the population along each dimension of the objective space. Inside each class, the individuals with the higher value of the crowding parameter obtain a better rank than those with a lower one.

At every generation, binary tournament selection, recombination, and mutation operators are used to create an offspring of the current population. The combination of the two is then sorted according to dominance first and then to crowding. The non-dominated individuals with lowest crowding factor are then used to update the population.

The parameters to be set are the size of the population, the number of generations, the crossover and mutation probability, $p_c$ and $p_m$, and distribution indexes for crossover and mutation, $\eta_c$ and $\eta_m$, respectively. Three different ratios between population size and number of generations were considered: 0.08, 0.33 and 0.75. The values $p_c$ and $p_m$ were set to 0.9 and 0.2 respectively and kept constant for all the tests. The values, 5, 10 and 20, were considered for $\eta_c$, while tests were run for values of $\eta_m$ equal to 5, 25 and 50.

\subsubsection{MOPSO}
MOPSO is an extension of Particle Swarm Optimization (PSO) to
multiobjective problems. Pareto dominance is introduced in the
selection criteria for the candidate solutions to update the
population. MOPSO features an external archive which stores all the
non-dominated solutions and at the same time is used to guide the
search process of the swarm. This is done by introducing the
possibility to direct the movement of a particle towards one of the
less crowded solutions in the archive. The solution space is subdivided into hypercubes through an adaptive grid. The
solutions in the external archive are thus reorganized in these
hypercubes. The algorithm keeps track of the crowding level of each
hypercube and promotes movements towards less crowded areas. In a
similar manner, it also gives priority to the insertion of new
non-dominated individuals in less crowded areas if the external
archive has already reached its predefined maximum size.  For MOPSO
three different ratios between population size and number of
generations were tested: 0.08 0.33 0.75. It was also tested with
three different numbers of subdivisions of the solution space: 10,
30, and 50. The inertia component in the motion of the particles was
set to 0.4 and the weights of the social and individualistic
components were se to 1

\subsubsection{PAES}
PAES is  a (1 + 1) Evolution Strategy with the addition of an external archive to store the current best approximation of the Pareto front. It adopts a population of only a single chromosome which, at every iteration, generates a mutated copy. The algorithm then preserves the non-dominated one between the parent and the candidate solution. If none of the two dominates the other, the algorithm then checks their dominance index with respect to the solutions in the archive. If also this comparison is inconclusive, then the algorithm selects the one which resides in the less crowded region of the objective space. To keep track of the crowding level of the objective space, the latter is subdivided in an n-dimensional grid. Every time a new solution is added to (or removed from) the archive, the crowding level of the corresponding grid cell is updated. PAES has two main parameters that need to be set, the number of subdivisions in the space grid and the mutation probability. Values of 1,2 and 4 were used for the former and 0.6 0.8 and 0.9 were used for the latter.

\subsubsection{MTS}
MTS is an algorithm based on pattern search. The algorithm first generates a population of uniformly distributed individuals. At every iteration, a local search is performed by a subset of individuals.  Three different search patterns are included in the local search: the first is a search along the direction of each decision variable with a fixed step length; the second is analogous but the search is limited to one fourth of all the possible search directions; the third one also searches along each direction but selects only solutions which are evenly spaced on each dimension within a predetermined upper bound and lower bound.
At each iteration and for each individual, the three search patterns are tested with few function evaluations to select the one which generates the best candidate solutions for the current individual. The selected one is then used to perform the local search which will update the individual itself. The step length along each direction, which defines the size of the search neighbourhood for each individual, is increased if the local search generated a non-dominated child and is decreased otherwise. When the neighbourhood size reaches a predetermined minimum value, it is reset to 40\% of the size of the global search space. The non-dominated candidate solutions are then used to update the best approximation of the global Pareto front. MTS was tested with a population size of 20, 40 and 80 individuals.

\subsection{Performance Metrics} Two metrics were defined to evaluate the performance of the tested
multiobjective optimizers:

\begin{equation}\label{eM1}
M_{spr}=\frac{1}{M_p}\sum_{i=1}^{M_p}\min_{j\in
N_p}100\Bigg\|\frac{\textbf{f}_j-\textbf{g}_i}{\textbf{g}_i}\Bigg\|
\end{equation}

\begin{equation}\label{eM2}
M_{conv}=\frac{1}{N_p}\sum_{i=1}^{N_p}\min_{j\in
M_p}100\Bigg\|\frac{\textbf{g}_j-\textbf{f}_i}{\textbf{g}_j}\Bigg\|
\end{equation}
where $M_p$ is the number of elements, with objective vector
$\mathbf{g}$, in the true global Pareto front and $N_p$ is the
number of elements, with objective vector $\mathbf{f}$, in the
Pareto front that a given algorithm is producing. Although similar,
the two metrics are measuring two different things: $M_{spr}$ is the
sum, over all the elements in the global Pareto front, of the
minimum distance of all the elements in the Pareto front $N_p$ from
the the $i^{th}$ element in the global Pareto front. $M_{conv}$,
instead, is the sum, over all the elements in the Pareto front
$N_p$, of the minimum distance of the elements in the global Pareto
front from the $i^{th}$ element in the Pareto front $N_p$.

Therefore, if $N_p$ is only a partial representation of the global
Pareto front but is a very accurate partial representation, then
metric $M_{spr}$ would give a high value and metric $M_{conv}$ a low
value. If both metrics are high then the Pareto front $N_p$ is
partial and poorly accurate. The index $M_{conv}$ is similar to the
mean Euclidean distance \cite{MOPSO:01}, although in $M_{conv}$ the
Euclidean distance is normalized with respect to the values of the
objective functions, while $M_{spr}$ is similar to the generational
distance \cite{Veldhuizen:98}, although even for $M_{spr}$ the
distances are normalized.

Given $n$ repeated runs of a given algorithm, we can define two
performance indexes: $p_{conv}=P(M_{conv} < tol_{conv})$ or the
probability that the index $M_{conv}$ achieves a value less than the
threshold $tol_{conv}$ and $p_{spr}=P(M_{spr}< tol_{spr})$ or the
probability that the index $M_{spr}$ achieves a value less than the
threshold $tol_{conv}$.

According to the theory developed in \cite{cit:vasileAAS2008,
minisci:09}, 200 runs are sufficient to have a 95\% confidence that
the true values of $p_{conv}$ and $p_{spr}$ are within a $\pm 5\%$
interval containing their estimated value.

Performance index (\ref{eM1}) and (\ref{eM2}) tend to uniformly
weigh every part of the front. This is not a problem for index
(\ref{eM1}) but if only a relatively small portion of the front is
missed the value of performance index (\ref{eM2}) might be only
marginally affected. For this reason, we slightly modified the
computation of the indexes by taking only the $M_P^*$ and $N_P^*$
solutions with a normalised distance in criteria space that was
higher than $10^{-3}$.

The global fronts used in the three space tests were built by taking
a selection of about 2000 equispaced, in criteria space,
nondominated solutions coming from all the 200 runs of all the
algorithms.

\subsection{Preliminary Test Cases}
For the preliminary tests, two sets of functions, taken from the
literature \cite{Deb:00,MOPSO:01}, were used and the performance of
MACS was compared to the results in \cite{Deb:00} and
\cite{MOPSO:01}. Therefore, for this set of tests, MTS was not
included in the comparison. The function used in this section can be
found in Table \ref{tab:multiobj}.

\begin{table}[ht]
\centering\caption{Multiobjective test
functions}\label{tab:multiobj}
\end{table}

The first three functions were taken from \cite{MOPSO:01}. Test cases
$Deb$ and $Scha$ are two examples of disconnected Pareto fronts,
$Deb2$ is an example of problem presenting multiple local Pareto
fronts, 60 in this two dimensional case.

The last three functions are instead taken from \cite{Deb:00}. Test
case $ZDT2$ has a concave Pareto front with a moderately
high-dimensional search space. Test case $ZDT6$ has a concave Pareto
front but because of the irregular nature of $f_1$ there is a strong
bias in the distribution of the solutions. Test case, \textit{ZDT4},
with dimension 10, is commonly recognized as one of the most
challenging problems since it has $21^9$ different local Pareto
fronts of which only one corresponds to the global Pareto-optimal
front.

As a preliminary proof of the effectiveness of MACS, the average
Euclidean distance of 500 uniformly spaced points on the true
optimal Pareto front from the solutions stored in $A_g$ by MACS was
computed and  compared to known results in the literature. MACS was
run 20 times to have a sample comparable to the one used for the
other algorithms. The global archive was limited to 200 elements to
be consistent with \cite{Deb:00}. The value of the crowding factor
$w_c$, the threshold $\rho_{tol}$ and the convergence $\rho_{min}$
were kept constant to 1e-5 in all the cases to provide good local
convergence.

To be consistent with \cite{MOPSO:01}, on $Deb$, $Scha$ and, $Deb2$,
MACS was run respectively for 4000, 1200 and 3200 function
evaluations. Only two agents were used for these lower dimensional
cases, with $f_e=1/2$. On test cases $ZDT2$, $ZDT4$ and $ZDT6$, MACS
was run for a maximum of 25000 function evaluations to be consistent
with \cite{Deb:00}, with three agents and $f_e=2/3$ for $ZDT2$ and
four agents and $f_e=3/4$ on $ZDT4$ and $ZDT6$.

The results on $Deb$, $Scha$ and, $Deb2$ can be found in
Table~\ref{tab:euc1}, while the results on $ZDT2$, $ZDT4$ and $ZDT6$
can be found in \ref{tab:euc2}.

On all the smaller dimensional cases MACS performs comparably to
MOPSO and better than PAES. It also performs than NSGA-II on \emph{Deb} and \emph{Deb2}. On \emph{Scha} MACS performs apparently worse than NSGA-II, although after inspection one can observe that all the elements of the global archive $A_g$ belong to the Pareto front but not uniformly distributed, hence the higher value of the Euclidean distance. On the higher dimensional
cases, MACS performs comparably to NSGA-II on ZDT2 but better than
all the others on ZDT4 and ZDT6. Note in particular the improved
performance on ZDT4.

\begin{table}[!hp]
\centering \caption{Comparison of the average Euclidean distances
between 500 uniformly space points on the optimal Pareto front for
various optimization algorithms: smaller dimension test
problems.}\label{tab:euc1}
\end{table}

\begin{table}[!hp]
\centering \caption{Comparison of the average Euclidean distances
between 500 uniformly space points on the optimal Pareto front for
various optimization algorithms: larger dimension test
problems.}\label{tab:euc2}
\end{table}

On the same six functions a different test was run to evaluate the
performance of different variants of MACS. For all variants, the
number of agents, $f_e$, $w_c$, $\rho_{tol}$ and $\rho_{min}$ was
set as before, but instead of the mean Euclidean distance, the
success rates $p_{conv}$ and $p_{spr}$ were measured for each
variant. The number of function evaluations for $ZDT2$, $ZDT4$,
$ZDT6$ and $Deb2$ is the same as before, while for $Scha$ and $Deb$
it was reduced respectively to 600 and 1000 function evaluations
given the good performance of MACS already for this number of function evaluations. Each run was repeated 200 times
to have good confidence in the values of  $p_{conv}$ and $p_{spr}$.

Four variants were tested and compared to the full version of MACS.
Variant MACS no local does not implement the individualistic moves
and the local archive, variant MACS $\rho=1$ has no adaptivity on
the neighborhood, its size is kept fixed to 1,variant MACS
$\rho=0.1$ has the size of the neighborhood fixed to 0.1, variant
MACS no attraction has the attraction mechanisms not active.

The result can be found in Table \ref{tab:macs_bench}. The values of
$tol_{conv}$ and $tol_{spr}$ are respectively 0.001 and 0.0035 for
$ZDT2$, 0.003 and 0.005 for $ZDT4$, 0.001 and 0.025 for $ZDT6$,
0.0012 and 0.035 for $Deb$, 0.0013 and 0.04 for $Scha$, 0.0015 and
0.0045 for $Deb2$. This thresholds were selected to highlight the
differences among the various variants.

The table shows that the adaptation mechanism is beneficial in some
cases although, in others, fixing the value of $\rho$ might be a
better choice. This depends on the problem and a general rule is
difficult to derive at present. Other adaptation mechanisms could
further improve the performance.

The use of individualistic actions coupled with a local archive is
instead fundamental, so is the use of the attraction mechanism.
Note, however, how the attraction mechanism penalizes the spreading
on biassed problems like $ZDT6$, this is expected as it accelerates
convergence.

\begin{table}[!hp]
\centering \caption{Comparison of different variants of
MACS.}\label{tab:macs_bench}
\end{table}

\subsection{Application to Space Trajectory Design}
\label{sec:app_space_traj}
In this section we present the application of MACS to three space trajectory problems: a
two-impulse orbit transfer from a Low Earth Orbit (LEO) to a
high-eccentricity Molniya-like orbit, a three-impulse transfer from
a LEO to Geostationary Earth Orbit (GEO) and a multi-gravity assist
transfer to Saturn equivalent to the transfer trajectory of the
Cassini mission. The first two cases are taken from the work of
Minisci et al.\cite{minisci:09}.

In the two-impulse case, the spacecraft departs at time $t_0$ from a
circular orbit around the Earth (the gravity constant is
$\mu_E=3.9860 · 10^5$ km$^3$s$^{-2}$) with radius $r_0=6721$ km and
at time $t_f$ is injected into an elliptical orbit with eccentricity
$e_T=0.667$ and semimajor axis $a_T=26 610$ km. The transfer arc is
computed as the solution of a Lambert's problem \cite{avanzini:08}
and the objective functions are the transfer time $T=t_f-t_0$ and
the sum of the two norms of the velocity variations at the beginning
and at the end of the transfer arc $\Delta v_{tot}$. The objectives
are functions of the solution vector $\mathbf{x}=[t_0\;t_f]^T\in
D\subset \mathbb{R}^2$ The search space $D$ is defined by the
following intervals $t_0\in[0\; 10.8]$, and $t_f\in[0.03 \; 10.8]$.

In the three-impulse case, the spacecraft departs at time $t_0$ from
a circular orbit around the Earth with radius $r_0=7000$ km and
after a transfer time $T=t_1+t_2$ is injected into a circular orbit
with radius $r_f=42000$. An intermediate manoeuvre is performed at
time $t_0+t_1$ and at position defined in polar coordinates by the
radius $r_1$ and the angle $\theta_1$. The objective functions are
the total transfer time $T$ and the sum of the three impulses
$\Delta v_{tot}$. The solution vector in this case is
$\mathbf{x}=[t_0,t_1,r_1,\theta_1,t_f]^T\in D\subset\mathbb{R}^5$.
The search space $D$ is defined by the following intervals
$t_0\in[0\; 1.62]$, $t_1\in[0.03\; 21.54]$, $r_1\in[7010\; 105410]$,
$\theta_1\in[0.01\; 2\pi-0.01]$, and $t_2\in[0.03\; 21.54]$.

The Cassini case consists of 5 transfer arcs connecting a departure
planet, the Earth, to the destination planet, Saturn, through a
sequence of swing-by's with the planets: Venus, Venus, Earth,
Jupiter. Each transfer arc is computed as the solution of a
Lambert's problem \cite{battin:99} given the departure time from
planet $P_i$ and the arrival time at planet $P_{i+1}$. The solution
of the Lambert's problems yields the required incoming and outgoing
velocities at each swing-by planet $v_{in}$ and $v_{rout}$. The
swing-by is modeled through a linked-conic approximation with
powered maneuvers\cite{becerra:04}, i.e., the mismatch between the
required outgoing velocity $v_{rout}$ and
 the achievable outgoing velocity $v_{aout}$ is compensated through a
$\Delta v$ maneuver at the pericenter of the gravity assist
hyperbola. The whole trajectory is completely defined by the
departure time $t_0$ and the transfer time for each leg $T_i$, with
$i=1,...,5$.
The normalized radius of the pericenter $r_{p,i}$ of each swing-by
hyperbola is derived a posteriori once each powered swing-by
manoeuvre is computed. Thus, a constraint on each pericenter radius
has to be introduced during the search for an optimal solution. In
order to take into account this constraint, one of the objective
functions is augmented with the weighted violation of the
constraints:
\begin{equation}\label{eq:evvejs_fobj}
    f(\mathbf{x})=\Delta v_0+\sum_{i=1}^{4}\Delta v_i + \Delta
v_f+\sum_{i=1}^{4}w_i(r_{p,i}-r_{pmin,i})^2
\end{equation}
for a solution vector $\mathbf{x}=[t_0,T_1,T_2,T_3,T_4,T_5]^T$.
The objective functions are, the total transfer time
$T=\sum_i^5 T_i$ and $f(\mathbf{x})$. The minimum normalized
pericenter radii are $r_{pmin,1}=1.0496$, $r_{pmin,2}=1.0496$,
$r_{pmin,3}=1.0627$ and $r_{pmin,4}=9.3925$. The search space $D$ is
defined by the following intervals: $t_0\in [-1000, 0]$MJD2000,
$T_1\in [30, 400]$d, $T_2\in [100, 470]$d, $T_3\in [30, 400]$d,
$T_4\in [400, 2000]$d, $T_5\in [1000, 6000]$d. The best known
solution for the single objective minimization of $f(\mathbf{x})$ is
$f_{best}=4.9307$ km/s, with $\mathbf{x}_{best}=[-789.753, 158.2993,
449.3859, 54.7060,$ $1024.5896, 4552.7054]^T$.

\subsubsection{Test Results}
For this second set of tests, each algorithm was run for the same
number of function evaluations. In particular, consistent with the
tests performed in the work of Minisci et al., \cite{minisci:09} we
used 2000 function evaluations for the two-impulse case and 30000
for the three-impulse case. For the Cassini case, instead, the
algorithms were run for 180000, 300000 and 600000 function
evaluations.

Note that the version of all the algorithms used in this second set
of tests is the one that is freely available online, written in
c/c++. We tried in all cases to stick to the available instructions
and recommendations by the author to avoid any bias in the
comparison.

The thresholds values for the two impulse cases was taken from
\cite{minisci:09} and is $tol_{conv}=0.1$, $tol_{spr}=2.5$. For the
three-impulse case instead we considered $tol_{conv}=5.0$,
$tol_{spr}=5.0$. For the Cassini case we used $tol_{conv}=0.75$,
$tol_{spr}=5$, instead. These values were selected after looking at
the dispersion of the results over 200 runs. Lower values would
result in a zero value of the performance indexes of all the
algorithms, which is not very significant for a comparison.

MACS was tuned on the three-impulse case. In particular, the
crowding factor $w_c$, the threshold $\rho_{tol}$ and the
convergence $\rho_{min}$ were kept constant to 1e-5, which is below
the required local convergence accuracy, while $f_e$ and $n_{pop}$
were changed. A value of 1e-5 is expected to provide good local
convergence and good density of the samples belonging to the Pareto
front. Table \ref{tab:macs_tuning} reports the value of performance
indexes $p_{conv}$ and $p_{spr}$ over 200 runs of MACS with
different settings. The index $p_{spr}$ and the index $p_{conv}$
have different, almost opposite, trends. However, it was decided to
select the setting that provides the best convergence,i.e.
$n_{pop}=15$ and $f_e=1/3$. This setting will be used for all the
tests in this paper.

On top of the complete algorithm, two variants of MACS were tested: one without individualistic moves and local archive, denoted as \emph{no local} in the tables, and one with no attraction towards the global archive $A_g$, denoted as \emph{no att} in the tables. Only these two variants are tested on these cases as they displayed the most significant impact in the previous standard test cases and more importantly were designed specifically to improve performances.

NSGA-II, PAES, MOPSO and MTS were tuned as well on the three-impulse
case. In particular, for NSGA-II the best result was obtained for
150 individuals and can be found in Table \ref{tab:NSGAII_3imp}.  A similar result could be obtained for MOPSO, see Table \ref{tab:MOPSO_3imp}. For
MTS only the population was changed while the number of individuals
performing local moves was kept constant to 5. The results of the tuning of MTS can be
found in Table \ref{tab:MTS_tuning}. For the tuning of PAES the
results can be found in Table \ref{tab:PAES_3imp}.

All the parameters tuned in the three impulse case were kept constant except for the population size of NSGA-II and MOPSO. The size of the population of NSGA-II and MOPSO was set to 100 and 40 respectively on the two impulse case and was increased with the number of function evaluations in the Cassini case. In particular for NSGA-II the following ratios between population size and number of function evaluations was used: 272/180000, 353/300000, 500/600000. For MOPSO the following ratios between population size and number of function evaluations was used: 224/180000, 447/300000, 665/600000.
This might not be the best way to set the population size for these two algorithms but it is the one that provided better performance in these tests.

Note that the size of the global archive for MACS was constrained to be lower than the size of the population of NSGA-II, in order to avoid any bias in the computation of $M_{spr}$.

The performance of all the algorithms on the Cassini case can be
found in Table \ref{tab:cassini} for a variable number of function
evaluations.

\begin{figure}[!hbt]
\begin{center}
 \caption{Three-impulse test case: a) Complete Pareto front, b) close-up of the Pareto fronts}
\end{center}
\end{figure}

\begin{figure}[!hbt]
\begin{center}
 \caption{Cassini test case: a) Complete Pareto front, b) close-up of the Pareto fronts}
\end{center}
\end{figure}

\begin{table}[thb!]
\centering \caption{Indexes $p_{conv}$ and $p_{spr}$ for
different settings of MACS} \label{tab:macs_tuning}
\end{table}

For the three-impulse case, MACS was able to identify an extended
Pareto front (see Fig.\ref{fig:3imp} and Fig. \ref{fig:3imp_close}
where all the non-dominated solutions from all the 200 runs are
compared to the global front), compared to the results in
\cite{minisci:09}. The gap in the Pareto front is probably due to a
limited spreading of the solutions in that region. Note the cusp due
the transition between the condition in which 2-impulse solutions
are optimal and the condition in which 3-impulse solutions are
optimal.

 Table
\ref{tab:bi-imp} summarizes the results of all the tested
algorithms on the two-impulse case. The average value of the
performance metrics is reported with, in brackets, the associated
variance over 200 runs. The two-impulse case is quite easy and all the
algorithms have no problems identifying the front. However, MACS
displays a better convergence than the other algorithms while the
spreading of MOPSO and MTS is superior to the one of MACS.

The three-impulse case is instead more problematic (see
Table~\ref{tab:tri-imp}). NSGA-II is
not able to converge to the upper-left part of the front and
therefore the convergence is 0 and the spreading is comparable to
the one of MACS. All the other algorithms perform poorly with a
value of almost 0 for the performance indexes. This is mainly due to
the fact that no one can identify the vertical part of the front.

Note that the long tail
identified by NSGA-II is actually dominated by two points belonging
to the global front (see \ref{fig:3imp_close}).

Table \ref{tab:cassini} reports the statistics for the Cassini
problem. On top of the performance of the three variants tested on the other two problems, the table reports also the result for 10
agents and $f_e=5$. For all numbers of function evaluations MACS has
better spreading than NSGA-II because NSGA-II converges to a local
Pareto front. Nonetheless NSGA-II displays a more regular behavior
and a better convergence for low number of function evaluations
although it never reaches the best front. The Pareto fronts are
represented in Fig. \ref{fig:Cassini_front} and Fig.
\ref{fig:Cassini_front_closeup} (even in this case the figures
represent all the non-dominated solutions coming from all the 200
runs). Note that, the minimum $f$ returned by MACS is the best known
solution for the single objective version of this problem (see
\cite{macs_jogo:08}). All the other algorithms perform quite poorly
on this case.

Of all the variants of MACS tested on these problems the full complete one is performing the best. As expected, in all three cases, removing the individualistic moves severely penalizes both convergence and spreading. It is interesting to note that removing the attraction towards the global front is also comparatively bad. On the two impulse case it does not impact the spreading but reduces the convergence, while on the Cassini case, it reduces mean and variance but the success rates are zero. The observable reason is that MACS converges slower but more uniformly to a local Pareto front.

Finally, it should be noted that mean and variance seem not to capture the actual performance of the algorithms. In particular they do not capture the ability to identify the whole Pareto front as the success rates instead do.

\begin{table}[thp!]
  \centering
    \caption{NSGAII tuning on the 3-impulse case}\label{tab:NSGAII_3imp}
\end{table}

\begin{table}[thp!]
  \centering
  \caption{PAES tuning on the 3-impulse case}\label{tab:PAES_3imp}
\end{table}

\begin{table}
  \centering
  \caption{MOPSO tuning on the 3-impulse case. }\label{tab:MOPSO_3imp}
\end{table}
\begin{table}[thb!]
  \centering
    \caption{MTS tuning on the 3-impulse}\label{tab:MTS_tuning}
\end{table}

\begin{table}[!hp]
\centering \caption{Summary of metrics $M_{conv}$ and $M_{spr}$, and
associated performance indexes $p_{conv}$ and $p_{spr}$, on the
three impulse test cases.}\label{tab:tri-imp}
\end{table}

\begin{table}[thp!]
\centering \caption{Metrics $M_{conv}$ and $M_{spr}$, and associated performance indexes $p_{conv}$ and $p_{spr}$, on the two impulse test cases.}\label{tab:bi-imp}
\end{table}

\begin{table}[thp!]
\centering \caption{Metrics $M_{conv}$ and $M_{spr}$, and associated
performance indexes $p_{conv}$ and $p_{spr}$, on the Cassini case.}\label{tab:cassini}
\end{table}

\section{Conclusions} \label{sec:conclusions}
In this paper we presented a hybrid evolutionary algorithm for
multiobjective optimization problems. The effectiveness of the
hybrid algorithm, implemented in a code called MACS, was
demonstrated at first on a set of standard problems and then its
performance was compared against NSGA-II, PAES, MOPSO and MTS on three space trajectory
design problems. The results are encouraging as, for the same
computational effort (measured in number of function evaluations,
MACS was converging more accurately than NSGA-II on the two-impulse
case and managed to find a previously undiscovered part of the
Pareto front of the three-impulse case. As a consequence, on the three-impulse case, MACS, has better performance metrics than the other algorithms. On the Cassini case NSGA-II
appears to converge better to some parts of the front although MACS
yielded solutions with better $f$ and identifies once more a part of
the front that NSGA-II cannot attain. PAES and MTS do not perform well on the Cassini case, while MOPSO converges well locally but, with the settings employed in this study,
yielded a very poor spreading.

From the experimental tests in this paper we can argue that the
following mechanisms seem to be particularly effective: the use of
individual local actions with a local archive as they allow the
individuals to move towards and within the Pareto set; the use of an
attraction mechanism as it accelerates convergence.

Finally it should be noted that all the algorithms tested in this
study use the Pareto dominance as selection criterion. Different
criteria, like the decomposition in scalar subproblems, can be
equality implemented in MACS, without disrupting its working
principles, and lead to different performance results.

\section*{Acknowledgements}
The authors would like to thank Dr.\ Edmondo Minisci and Dr.\ Guilio Avanzini for the two and three impulse cases and the helpful advice.

\bibliographystyle{unsrt}
\bibliography{epicmoo_cec}

\begin{thebibliography}{10}

\bibitem{coverstone:00}
V.~Coverstone-Caroll, J.W. Hartmann, and W.M. Mason.
\newblock Optimal multi-objective low-thrust spacecraft trajectories.
\newblock {\em Computer Methods in Applied Mechanics and Engineering},
  186:387--402, 2000.

\bibitem{dachwald:04}
B.~Dachwald.
\newblock Optimization of interplanetary solar sailcraft trajectories using
  evolutionary neurocontrol.
\newblock {\em Journal of Guidance, and Dynamics}, February 2004.

\bibitem{lee:05}
S.~Lee, P.~von Allmen, W.~Fink, A.E. Petropoulos, and R.J. Terrile.
\newblock Multi-objective evolutionary algorithms for low-thrust orbit transfer
  optimization.
\newblock In {\em Proceedings of the Genetic and Evolutionary Computation
  Conference (GECCO 2005)}, Washington DC, USA, June 25--29 2005.

\bibitem{schuetze:08c}
O.~Sch\"{u}tze, M.~M.~Vasile, and C.A.~Coello Coello.
\newblock Approximate solutions in space mission design.
\newblock In {\em Parallel Problem Solving from Nature (PPSN 2008)}, Dortmund,
  Germany, September 13--17 2008.

\bibitem{schuetze:09}
O.~Sch\"{u}tze, M.~Vasile, O.~Junge, M.~Dellnitz, and D.~Izzo.
\newblock Designing optimal low-thrust gravity-assist trajectories using space
  pruning and a multi-objective approach.
\newblock {\em Engineering Optimization}, 41(2), February 2009.

\bibitem{delnitz:09}
M.~Dellnitz, S.~Ober-Blöbaum, M.~Post, O.~Schütze, and Bianca Thiere.
\newblock A multi-objective approach to the design of low thrust space
  trajectories using optimal control.
\newblock {\em Celestial Mechanics and Dynamical Astronomy}, 105(1), February
  2009.

\bibitem{minisci:09}
E.~Minisci and G.~Avanzini.
\newblock Optimisation of orbit transfer manoeuvres as a test benchmark for
  evolutionary algorithms.
\newblock In {\em Proceedings of the 2009 IEEE Congress on Evolutionary
  Computation (CEC2009)}, Trondheim, Norway, May 2009.

\bibitem{vasile_cec:07}
M.~Vasile.
\newblock A behavioral-based meta-heuristic for robust global trajectory
  optimization.
\newblock In {\em Proceedings of the 2007 IEEE Congress on Evolutionary
  Computation (CEC2007)}, Singapore, September 2007.

\bibitem{macs_jogo:08}
M.~Vasile and M.~Locatelli.
\newblock A hybrid multiagent approach for global trajectory optimization.
\newblock {\em Journal of Global Optimization}, 44(4):461--479, August. 2009.

\bibitem{schuetze_hcs:08}
Oliver Sch\"utze, Adriana Lara, Gustavo Sanchez, and Carlos A.~Coello Coello.
\newblock {HCS}: A new local search strategy for memetic multi-objective
  evolutionary algorithms.
\newblock {IEEE} {T}ransactions on {E}volutionary {C}omputation (to appear),
  2010.

\bibitem{MOEAD07}
Q.~Zhang and H.~Li.
\newblock Moea/d: A multiobjective evolutionary algorithm based on
  decomposition.
\newblock {\em IEEE Transactions on Evolutionary Computation}, 11(6), November.

\bibitem{kenneth:05}
K.V. Price, R.M. Storn, and J.A. Lampinen.
\newblock {\em Differential Evolution. A Practical Approach to Global
  Optimization}.
\newblock Natural Computing Series. Springer, 2005.

\bibitem{MACS:10}
M.~Vasile and F.~Zuiani.
\newblock A hybrid multiobjective optimization algorithm applied to space
  trajectory optimization.
\newblock In {\em Proceedings of the IEEE International Conference on
  Evolutionary Computation}, Barcelona, Spain, July 2010.

\bibitem{Deb:00}
K.~A. Deb, A.~Pratap, and T.~Meyarivan.
\newblock Fast elitist multi-objective genetic algorithm: Nga-ii.
\newblock Kangal report no. 200001, KanGAL, 2000.

\bibitem{MOPSO:01}
C.~Coello and M.~Lechuga.
\newblock Mopso: A proposal for multiple objective particle swarm optimization.
\newblock Technical report evocinv-01-2001., CINVESTAV. Instituto Politecnico
  Nacional. Col. San Pedro Zacatenco. Mexico., 2001.

\bibitem{PAES:99}
J.D. Knowles and D.W. Corne.
\newblock The pareto archived evolution strategy : A new baseline algorithm for
  pareto multiobjective optimisation.
\newblock In {\em Proceedings of the IEEE International Conference on
  Evolutionary Computation}, Washington DC, US, 1999.

\bibitem{MTS:07}
Lin-Yu Tseng and Chun Chen.
\newblock Multiple trajectory search for multiobjective optimization.
\newblock In {\em Proceedings of the IEEE International Conference on
  Evolutionary Computation}, Singapore, 25-28 September 2007.

\bibitem{Veldhuizen:98}
D.A.~Van Veldhuizen and G.B. Lamont.
\newblock Evolutionary computation and convergence to a pareto front.
\newblock In {\em Late Breaking papers at the Genetic Programming}, 1998.

\bibitem{cit:vasileAAS2008}
M.~Vasile, E.~Minisci, and M.~Locatelli.
\newblock On testing global optimization algorithms for space trajectory
  design.
\newblock In {\em AIAA/AAS Astrodynamic Specialists Conference}, Honolulu,
  Hawaii, USA, Aug 2008.

\bibitem{avanzini:08}
G.~Avanzini.
\newblock A simple lambert algorithm.
\newblock {\em Journal of Guidance, Control, and Dynamics}, 31(6):1587--1594,
  Nov.--Dec. 2008.

\bibitem{battin:99}
R.~Battin.
\newblock {\em An Introduction to the Mathematics and Methods of
  Astrodynamics}.
\newblock AIAA, 1999.

\bibitem{becerra:04}
D.~R. Myatt, V.M. Becerra, S.J. Nasuto, and J.M. Bishop.
\newblock Global optimization tools for mission analysis and design.
\newblock Final rept. esa ariadna itt ao4532/18138/04/nl/mv,call03/4101,
  ESA/ESTEC, 2004.

\end{thebibliography}

\clearpage
\appendix
\section*{nomenclature}
\begin{longtable}[t]{p{1,4cm}p{11cm}}
$A_g$ & global archive\\
$A_l$ & local archive\\
$a_T$ & semimajor axis\\
$D$& search space\\
$e_T$ & eccentricity\\
$f$& cost function\\
$f_e$& fraction of the population doing local moves\\
$I_d$ & dominance index\\
$M_{conv}$& convergence metrics\\
$M_{spr}$& spreading metrics\\
$N_\rho$ & neighborhood of solution $\mathbf{x}$\\
$N_e$ & maximum number of allowed function evaluations \\
$n_{eval}$ & number of function evaluations \\
$n_{pop}$& population size\\
$P_i$ & i-th planet\\
$P_k$& population at generation $k$\\
$p_{conv}$ & percentage of success on convergence\\
$p_{spr}$ & percentage of success on spreading\\
$r$ & random number \\
$r_p$& pericentre radius\\
$r_{pmin}$& minimum pericentre radius\\
$S$& selection function\\
$s$ & resource index \\
$T$& transfer time \\
$t_0$& departure time\\
$t_i$& manoeuvre time\\
$t_f$& final time\\
$U$ & uniform distribution \\
$\mathbf{u}$& variation of the solution $\mathbf{x}$\\
$X$ & Pareto optimal set\\
$\mathbf{x}$& solution vector\\
$\mathbf{y}$ & mutate individual \\
$w_c$ & tolerance on the maximum allowable crowding \\
& \\
& \textit{Greek symbols} \\
$\Delta v$& variation of velocity\\
$\rho$ & size of the neighborhood $N_\rho$\\
$\mu_E$ & gravity constant\\
$\theta_i$& true anomaly of manoeuvre $i$ \\
\end{longtable}
\clearpage
\section*{List of Tables}
\setcounter{table}{0}

\begin{table}[!hp]
\centering\caption{Multiobjective test functions}\label{tab:multiobj}
\begin{tabular}{cl}
\toprule
\it Scha &  $f_2=(x-5)^2 $ \\
$x\in[-5,10]$ & $f_1=\left\{\begin{array}{lll}
                                            -x &if &x \le 1 \\
                                            -2+x &if& 1<x<3 \\
                                            4-x &if &3<x\le4\\
                                            -4+x &if& x>4
                                          \end{array}\right. $ \\
\hline
\it Deb &  $f_1=x_1$\\
$x_1,x_2\in[0,1]$
&$f_2=(1+10x_2) \Big[ 1-\left(\frac{x_1}{1+10x_2}\right)^\alpha-$\\
$\alpha=2$;  $q=4$ &$\frac{x_1}{1+10x_2}\sin(2\pi qx_1)\Big]$ \\
\hline
\it Deb2 &  $f_1=x_1$\\
$x_1\in[0,1]$&$f_2=g(x_1,x_2)h(x_1,x_2); g(x_1,x_2)=11+x_2^2-10\cos(2\pi x_2)$\\
$x_2\in[-30,30]$ & $h(x_1,x_2)=\Big \{\begin{array}{c}
                                     1-\sqrt{\frac{f_1}{g}}\;\; \mathrm{if} f_1\leq g \\
                                     0\;\; \mathrm{otherwise}
                                   \end{array}$ \\
\hline \it ZDT2 & $g=1+\frac{9}{n-1}\sum_{i=2}^n x_i$\\
$ x_i\in[0,1];$ & $h=1-(\frac{f_1}{g})^2$  \\
$i=1,\ldots,n$ & $f_1=x_1; f_2=gh$\\
$n=30$ \\
\hline \it ZDT4 & $g=1+10(n-1)+\sum_{i=2}^n [x_i^2-10 \cos(2\pi q
x_i)];$\\
$ x_1\in[0,1];$ & $h=1-\sqrt{\frac{f_1}{g}}$  \\
$x_i\in[-5,5];$ & $f_1=x_1; f_2=gh$\\
$i=2,\ldots,n$&   \\
$n=10$ \\
\hline \it ZDT6 & $g=1+9\sqrt[4]{\frac{\sum_{i=2}^n x_i}{n-1}}$\\
$ x_i\in[0,1];$ & $h=1-(\frac{f_1}{g})^2$  \\
$i=1,\ldots,n$ & $f_1=1-\exp(-4 x_1)\sin^6(6\pi x_1); f_2=gh$\\
$n=10$ \\ \bottomrule
\end{tabular}
\end{table}

\begin{table}[!hp]
\centering \caption{Comparison of the average Euclidean distances
between 500 uniformly space points on the optimal Pareto front for
various optimization algorithms: smaller dimension test problems.}\label{tab:euc1}
\begin{tabular}{lccc}
\toprule
Approach  & Deb2       & Scha        & Deb \\
\midrule
MACS      & 1.542e-3   &  3.257e-3  &  7.379e-4 \\
          & (5.19e-4)  &  (5.61e-4)   &  (6.36e-5)\\
NSGA-II   & 0.094644   &  0.001594   &  0.002536  \\
          & (0.117608) & (0.000122)  &  (0.000138) \\
PAES      & 0.259664   &  0.070003   &  0.002881   \\
          &  (0.573286)&  (0.158081) &  (0.00213) \\
MOPSO     & 0.0011611  &  0.00147396 &  0.002057  \\
          & (0.0007205)&  (0.00020178)&  (0.000286)\\
\bottomrule
\end{tabular}
\end{table}

\begin{table}[!hp]
\centering \caption{Comparison of the average Euclidean distances
between 500 uniformly space points on the optimal Pareto front for
various optimization algorithms: larger dimension test problems.}\label{tab:euc2}
\begin{tabular}{lccc}
\toprule
Approach  &   ZDT2       & ZDT4         & ZDT6 \\
\midrule
MACS      & 9.0896e-4  & 0.0061   &   0.0026 \\
          & (4.0862e-5)& (0.0133)  &  (0.0053)  \\
NSGA-II   & 0.000824   & 0.513053   & 0.296564\\
          &  ($<$1e-5)   & (0.118460) & (0.013135)\\
PAES      &  0.126276  & 0.854816   & 0.085469\\
          &  (0.036877)&  (0.527238)& (0.006644)\\
\bottomrule
\end{tabular}
\end{table}

\begin{table}[!hp]
\centering \caption{Comparison of different versions of MACS.}\label{tab:macs_bench}
\begin{tabular}{lccccccc}
\toprule
Approach           & Metric & ZDT2      &    ZDT4   & ZDT6     & Scha & Deb   & Deb2\\
\midrule
MACS               &$p_{conv}$& 83.5\%    & 75\%    &  77\%  &73\%  &70.5\% & 60\%\\
                   &$p_{spr}$ & 22.5\%    & 28\%    &  58.5\%  &38.5\%& 83\%&67.5 \%\\
MACS       &$p_{conv}$&  14\%   & 0\%     &  45\%  &0.5\% & 72.5\%&11\%\\
no local                   &$p_{spr}$ &  1\%    & 0\%     &  34\%  & 0\%  &  4\%  &15\%\\
MACS       &$p_{conv}$&  84\%   & 22\%    &  78\%  & 37\% &  92\% &21\%\\
$\rho=1$                   &$p_{spr}$ &  22\%   & 7\%     &  63\%  & 0\%  &  54\% &38\%\\
MACS     &$p_{conv}$& 56\%    & 42\%    &  57\%  & 78\% &  85\%     &42\%\\
$\rho=0.1$                   &$p_{spr}$ &  5\%    & 15\%    &  61\%  & 88\% &  94.5\%     &74\%\\
MACS  &$p_{conv}$&   21\% &  0.5\% &  14.5\%  & 0.5\% & 88.5\%&0\%\\
no attraction                   &$p_{spr}$ &  0\%   &  0.5\% & 78.5\%   &  0\%  &  0\%  &0\%\\
\bottomrule
\end{tabular}
\end{table}


\begin{table}[!hp]
\centering \caption{Indexes $p_{conv}$ and $p_{spr}$ for different
settings of MACS on the 3-impulse case}\label{tab:macs_tuning}
\begin{tabular}{l ccc}  
\toprule
  $p_{conv}$ & $n_{pop}=5$ & $n_{pop}=10$ & $n_{pop}=15$  \\
\midrule
 $f_e=1/3$ & 45.5\%  & 55.5\%  & 61.0\%   \\
 $f_e=1/2$ & 48.0\%  & 51.0\%  & 55.5\%   \\
 $f_e=2/3$ & 45.0\%  & 52.5\%  & 43.0\%  \\
\midrule
  $p_{spr}$ & $n_{pop}=5$ & $n_{pop}=10$ & $n_{pop}=15$  \\
\hline
 $f_e=1/3$ & 68.5\%  & 62.0\%  & 56.0\%   \\
 $f_e=1/2$ & 65.0\%  & 57.0\%  & 46.0\%  \\
 $f_e=2/3$ & 67.5\%  & 51.0\%  & 36.5\%   \\
\bottomrule
\end{tabular}
\end{table}

\begin{table}
  \centering
    \caption{NSGAII tuning on the 3-impulse case}\label{tab:NSGAII_3imp}
  \begin{tabular}{ccccccccc}
\toprule
 Mean\; $M_{conv}$ &    & & &Var\; $M_{conv}$ & & &  \\
    \hline
$\eta_c$/ $\eta_m$  &5      &25    &50  & $\eta_c$/ $\eta_m$    &5  &25 &50\\
    \hline
5   &36.1   &38.3   &43.0       &5  &201.0  &202.0  &185.0\\
10  &32.3   &39.4   &40.6       &10 &182.0  &172.0  &182.0\\
20  &31.7   &39.6   &42.5       &20 &175.0  &183.0  &169.0\\
    \hline
 Mean\; $M_{spr}$ & & & &Var\; $M_{spr}$ & & &  \\
    \hline
$\eta_c$/ $\eta_m$  &5  &25 &50     &$\eta_c$/ $\eta_m$ &5  &25 &50\\
    \hline
5   &6.77   &7.24   &8.08       &5  &9.97   &9.47   &7.25\\
10  &5.91   &7.50   &7.81       &10 &9.74   &8.68   &8.34\\
20  &5.78   &7.50   &8.16       &20 &9.75   &8.53   &8.04\\
    \hline
$p_{conv}$          &   &   & &$p_{spr}$ & &            \\
    \hline
$\eta_c$/ $\eta_m$  &5  &25 &50     &$\eta_c$/ $\eta_m$ &5  &25 &50\\
    \hline
5   &0.0\%  &0.0\%  &0.0\%      &5  &44.8\% &37.7\% &23.4\%\\
10  &0.0\%  &0.0\%  &0.0\%      &10 &57.8\% &33.1\% &29.2\%\\
20  &0.0\%  &0.0\%  &0.0\%      &20 &61.0\% &32.5\% &24.7\%\\
\bottomrule
  \end{tabular}
\end{table}

\begin{table}
  \centering
  \caption{MOPSO tuning on the 3-impulse case. }\label{tab:MOPSO_3imp}
  \begin{tabular}{ccccccccc}
\toprule
 Mean\; $M_{conv}$ &    & & &Var\; $M_{conv}$ & & &  \\
    \hline
Particles/Subdivisions  &10     &30    &50  & Particles/Subdivisions    &10 &30  &50\\
    \hline
50              &59.1   &50.2   &41.5       &50 &1080.0 &1010.2 &713.3\\
100             &50.0   &43.3   &41.3       &100&591.0  &721.1  &778.1\\
150             &47.8   &41.4   &39.4       &150&562.1  &550.2  &608.2\\
    \hline
 Mean\; $M_{spr}$ & & & &Var\; $M_{spr}$ & & &  \\
    \hline
Particles/Subdivisions  &10 &30 &50     &Particles/Subdivisions &10 &30 &50\\
    \hline
50  &16.1   &13.5   &12.3       &50     &41.3   &37.3   &24.0\\
100 &14.7   &12.2   &11.6       &100    &32.8   &25.8   &24.5\\
150 &14.8   &11.9   &11.4       &150    &30.0   &22.2   &22.5\\
    \hline
$p_{conv}$          &   &   & &$p_{spr}$ & &            \\
    \hline
Particles/Subdivisions  &10 &30 &50     &Particles/Subdivisions &10 &30 &50\\
    \hline
50  &0\%    &0\%    &0\%        &50     &0.5\%  &2.5\%  &3.5\%\\
100 &0\%    &0\%    &0\%        &100    &0.5\%  &4.5\% &4\%\\
150 &0\%    &0\%    &0\%        &150    &0.5\%  &2\% &4\%\\
\bottomrule
  \end{tabular}
\end{table}

\begin{table}
  \centering
  \caption{PAES tuning on the 3-impulse case}\label{tab:PAES_3imp}
  \begin{tabular}{ccccccccc}
\toprule
 Mean\; $M_{conv}$ &    & & &Var\; $M_{conv}$ & & &  \\
    \hline
Subdivisions/Mutation   &0.6        &0.8       &0.9 & Subdivisions/Mutation &0.6     &0.8    &0.9\\
    \hline
1               &53.7   &70.6   &70.2       &1  &525.0  &275.0  &297.0\\
2               &52.8   &70.2   &70.0       &2  &479.0  &266.0  &305.0\\
4               &53.0   &70.2   &70.1       &4  &453.0  &266.0  &311.0\\
    \hline
 Mean\; $M_{spr}$ & & & &Var\; $M_{spr}$ & & &  \\
    \hline
Subdivisions/Mutation   &0.6        &0.8       &0.9 &Subdivisions/Mutation  &0.6        &0.8       &0.9\\
    \hline
1   &14.2   &27.7   &36.7       &1  &20.0   &14.3   &17.3\\
2   &13.6   &27.6   &36.6       &2  &17.5   &14.6   &16.8\\
4   &13.8   &27.7   &36.6       &4  &17.2   &15.8   &17.0\\
    \hline
$p_{conv}$          &   &   & &$p_{spr}$ & &            \\
    \hline
Subdivisions/Mutation   &0.6    &0.8    &0.9        &Subdivisions/Mutation  &0.6     &0.8    &0.9\\
    \hline
1   &0.0\%  &0.0\%  &0.0\%      &1  &0.0\%  &0.0\%  &0.0\%\\
2   &0.0\%  &0.0\%  &0.0\%      &2  &0.0\%  &0.0\%  &0.0\%\\
4   &0.0\%  &0.0\%  &0.0\%      &4  &0.0\%  &0.0\%  &0.0\%\\
\bottomrule
  \end{tabular}
\end{table}

\begin{table}
  \centering
    \caption{MTS tuning on the 3-impulse}\label{tab:MTS_tuning}
\begin{tabular}{ccccccccc}
\toprule
3imp    &Population &20 &40 &80 \\
  \hline
$M_{conv}$  &Mean   &17.8   &22.6   &23.6\\
            &Var    &97.6   &87.8   &73.2\\
$M_{spr}$       &Mean   &12.6   &19.9   &18.4\\
            &Var    &34.7   &26.2   &18.6\\
            &$p_{conv}$ &1.0\%  &0.0\%  &0.0\%\\
            &$p_{spr}$  &0.5\%  &0.0\%  &0.0\%\\
\bottomrule
\end{tabular}
\end{table}

\begin{table}[!hp]
\centering \caption{Summary of metrics $M_{conv}$ and $M_{spr}$, and
associated performance indexes $p_{conv}$ and $p_{spr}$, on the
three impulse test cases.}\label{tab:tri-imp}
\begin{tabular}{@{} c ccccccc@{}}
\toprule
 Metric & MACS & MACS     & MACS  & NSGA-II  & PAES & MOPSO & MTS \\
        &      & no local &  no att     & &  &  & \\
\midrule
$M_{conv}$ & 5.53 & 7.58   & 154.7 & 31.7  &53.0 &39.4 & 17.8 \\
           &(15.1)&(26.3)  & (235.0) &(175.0)&(453.0)&(608.1)&(97.6)\\
$M_{spr}$  & 5.25 & 6.03   & 9.16 & 5.78   &13.8  &11.4  & 12.6 \\
           &(3.73)&(3.95)  & (2.07) &(9.75)&(17.2)&(22.5)&(34.7)\\
$p_{conv}$ & 61.0\%& 40.5\%  & 0.0\%        & 0.0\%        &0.0\%       &0\%         & 1.0\%\\
$p_{spr}$  & 56.0\%& 36\%  & 0.0\%        & 61.0\%       &0.0\%       &4.0\%       & 0.5\%\\
\bottomrule
\end{tabular}
\end{table}

\begin{table}[!hp]
\centering \caption{Metrics $M_{conv}$ and $M_{spr}$, and associated
performance indexes $p_{conv}$ and $p_{spr}$, on the two impulse
test cases.}\label{tab:bi-imp}
\begin{tabular}{@{} c ccccccc@{}}
\toprule
 Metric & MACS &    MACS  &     MACS  & NSGA-II  & PAES & MOPSO & MTS \\
        &      & no local &     no att      &&  &  & \\
\midrule
$M_{conv}$ & 0.0077  & 0.0039  &0.0534& 0.283   &0.198  &0.378  & 0.151 \\
           &(1e-3)   &(1.4e-2) &(9.5e-3)&(4.35e-3)&(0.332)&(0.0636)&(0.083)\\
$M_{spr}$  & 2.89    & 6.87    &3.08& 2.47  &332.0 &2.11  & 1.95 \\
           &(0.49)   &(7.36)   &0.943&(0.119)&(2.61e4)&(2.65)&(1.41)\\
$p_{conv}$ & 98.5\%  & 91.5\%  &84\%& 0\%  &75.5\%&9\%&57.5\%\\
$p_{spr}$  & 29.5\%  & 0\%     &24\%& 62.5\%  &0.5\%&94.5\%& 85.5\% \\
\bottomrule
\end{tabular}
\end{table}

\begin{table}[!hp]
\centering \caption{Metrics $M_{conv}$ and $M_{spr}$, and associated
performance indexes $p_{conv}$ and $p_{spr}$, on the Cassini case.}\label{tab:cassini}
\begin{tabular}{@{}l c ccc@{}}
\toprule
Approach & Metric & 180k & 300k & 600k  \\
\midrule
MACS         & $M_{conv}$ & 6.50 (229.1)& 4.48 (107.1) & 3.91 (62.6)\\
 5/15                  & $M_{spr}$  & 12.7 (126.0) & 11.1 (83.1)  & 8.64 (35.5)\\
                   & $p_{conv}$ & 6\%         & 14\%         & 21.5\%\\
                   & $p_{spr}$  & 27.5\%      & 31\%       & 41\%\\
\midrule
MACS         & $M_{conv}$ & 6.74 (217.9) & 5.56 (136.4)     & 3.14 (50.1)\\
      5/10             & $M_{spr}$  & 12.1 (83.0) & 10.3 (63.7)     & 8.11 (30.0)\\
                   & $p_{conv}$ & 8\%       & 10\%           & 25.5\%\\
                   & $p_{spr}$  & 26.0\%      & 31.5\%          & 45.5\%\\
\midrule
MACS               & $M_{conv}$ & 13.5 (436.2) & 10.2 (350.1)& 7.86 (190.2)\\
no local   & $M_{spr}$  & 31.1 (274.3) & 27.9 (278.9)& 21.9 (226.7)\\
                   & $p_{conv}$ & 1.0\%        & 1.5\%       & 2.5\%\\
                   & $p_{spr}$  & 1.0\%        & 2.5\%       & 6\%\\
\midrule
MACS               & $M_{conv}$ & 2.62 (1.46) & 2.2 (0.88)& 1.82 (0.478)\\
no att             & $M_{spr}$  & 27.8 (83.2) & 22.9 (58.0)& 18.2 (28.0)\\
                   & $p_{conv}$ & 0.0\%        & 0.0\%       & 1.0\%\\
                   & $p_{spr}$  & 0.0\%        & 0.0\%       & 0.0\%\\
\midrule

NSGA-II & $M_{conv}$ & 2.43 (18.0) & 1.99 (16.8)  & 1.24 (1.62)  \\
        & $M_{spr}$  & 11.6 (71.4) & 11.0 (47.5)  & 8.78 (28.2)  \\
        & $p_{conv}$ & 17.5\%      & 24.0\%       & 29.0\%  \\
        & $p_{spr}$  & 15.5\%      & 12.5\%       & 25.0\%  \\
\midrule
MOPSO & $M_{conv}$   & 2.62 (7.33) & 2.4 (2.57)  & 2.14 (0.94) \\
        & $M_{spr}$  & 28.0 (308.3) & 24.6 (260.4)  & 21.8 (231.3)  \\
        & $p_{conv}$ & 0.5\%      & 1.0\%       & 1.0\%  \\
        & $p_{spr}$  & 0.0\%      & 0.0\%       & 0.5\%  \\
\midrule
PAES    & $M_{conv}$ & 24.0 (54.5) & 19.8 (32.9)  & 15.2 (16.6)\\
        & $M_{spr}$  & 30.1 (47.3) & 26.0 (33.9)  & 21.4 (19.5)\\
        & $p_{conv}$ & 0.0\%      & 0.0\%       & 0.0\%  \\
        & $p_{spr}$  & 0.0\%      & 0.0\%       & 0.0\%  \\
\midrule
MTS & $M_{conv}$ & 3.71 (1.53) & 3.39 (1.67)  & 3.02 (1.69)  \\
    & $M_{spr} $ & 18.1 (18.2)& 15.6 (13.4)  & 13.1 (8.46)\\
        & $p_{conv}$ & 0.0\%  & 0.0\%       & 0.0\%  \\
        & $p_{spr}$  & 0.0\%  & 0.0\%       & 0.0\%  \\
\bottomrule
\end{tabular}
\end{table}

\clearpage
\section*{List of Figures}

\setcounter{figure}{0}
\begin{figure}
\begin{center}
  \subfigure[\label{fig:local}]{\includegraphics[width=7.5cm]{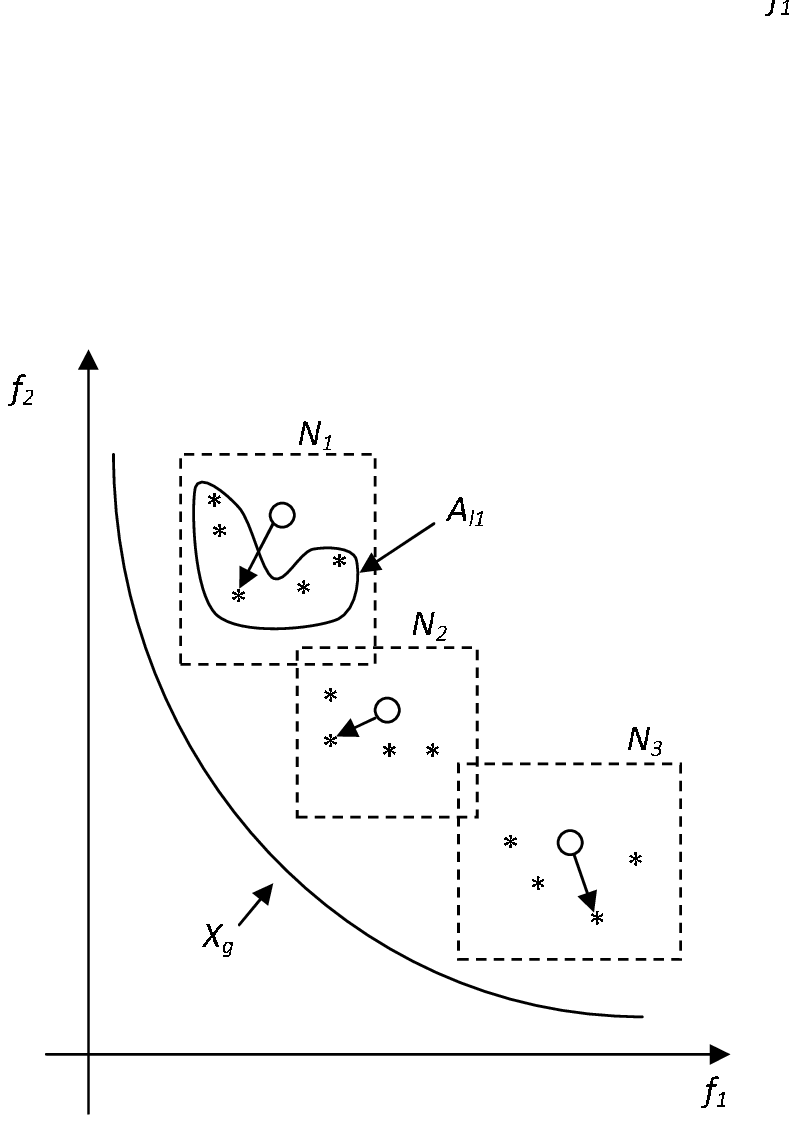}}
  \subfigure[\label{fig:global}]{\includegraphics[width=7.5cm]{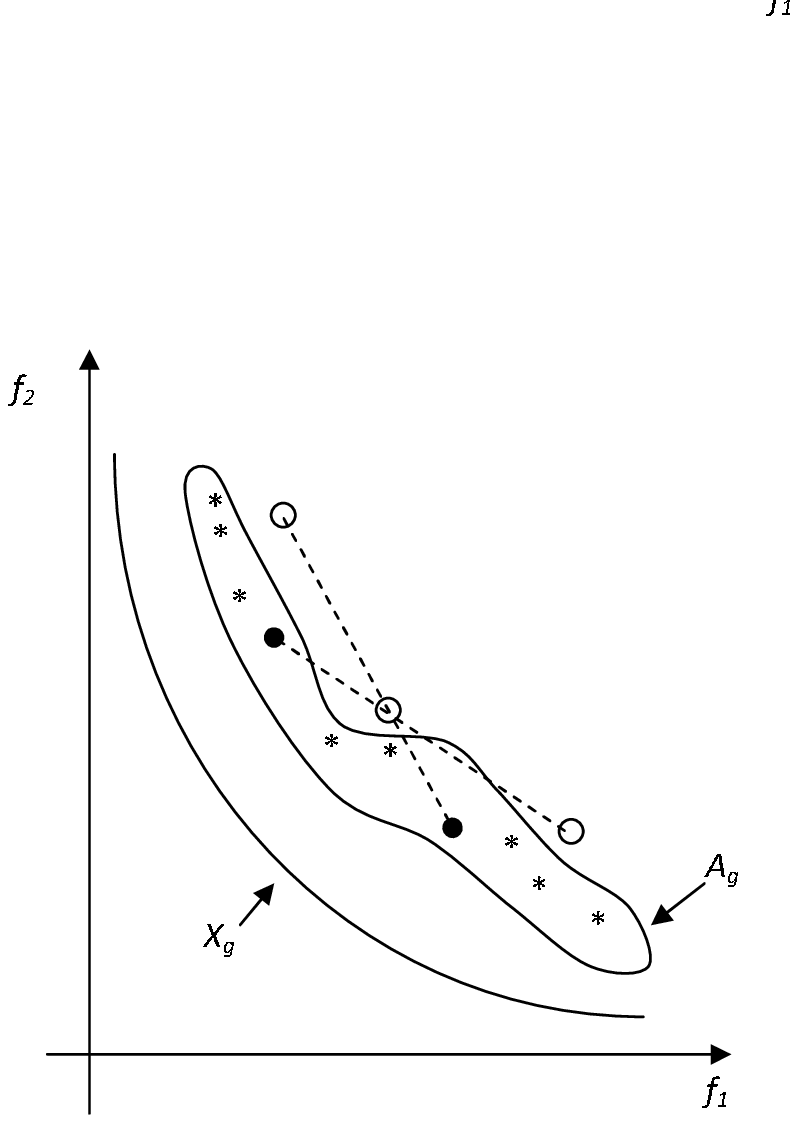}}
  \caption{Illustration of the a) local moves and archive and b) global moves and archive.}\label{fig:local_global}
  \end{center}
\end{figure}

\begin{figure}[!h]
\begin{center}
\subfigure[\label{fig:3imp}]{\includegraphics[width=7.5cm]{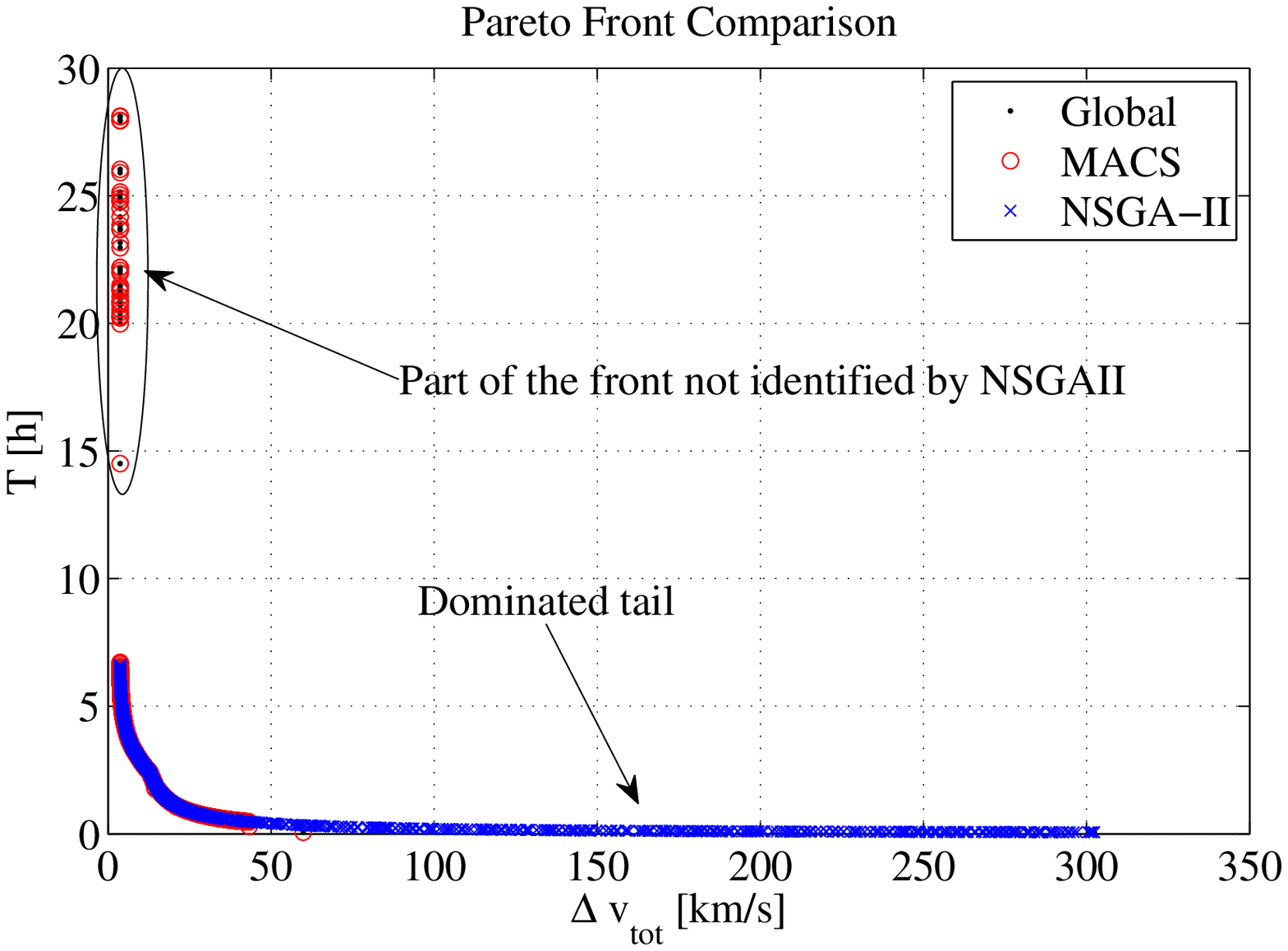}}
\subfigure[\label{fig:3imp_close}]{\includegraphics[width=7.5cm]{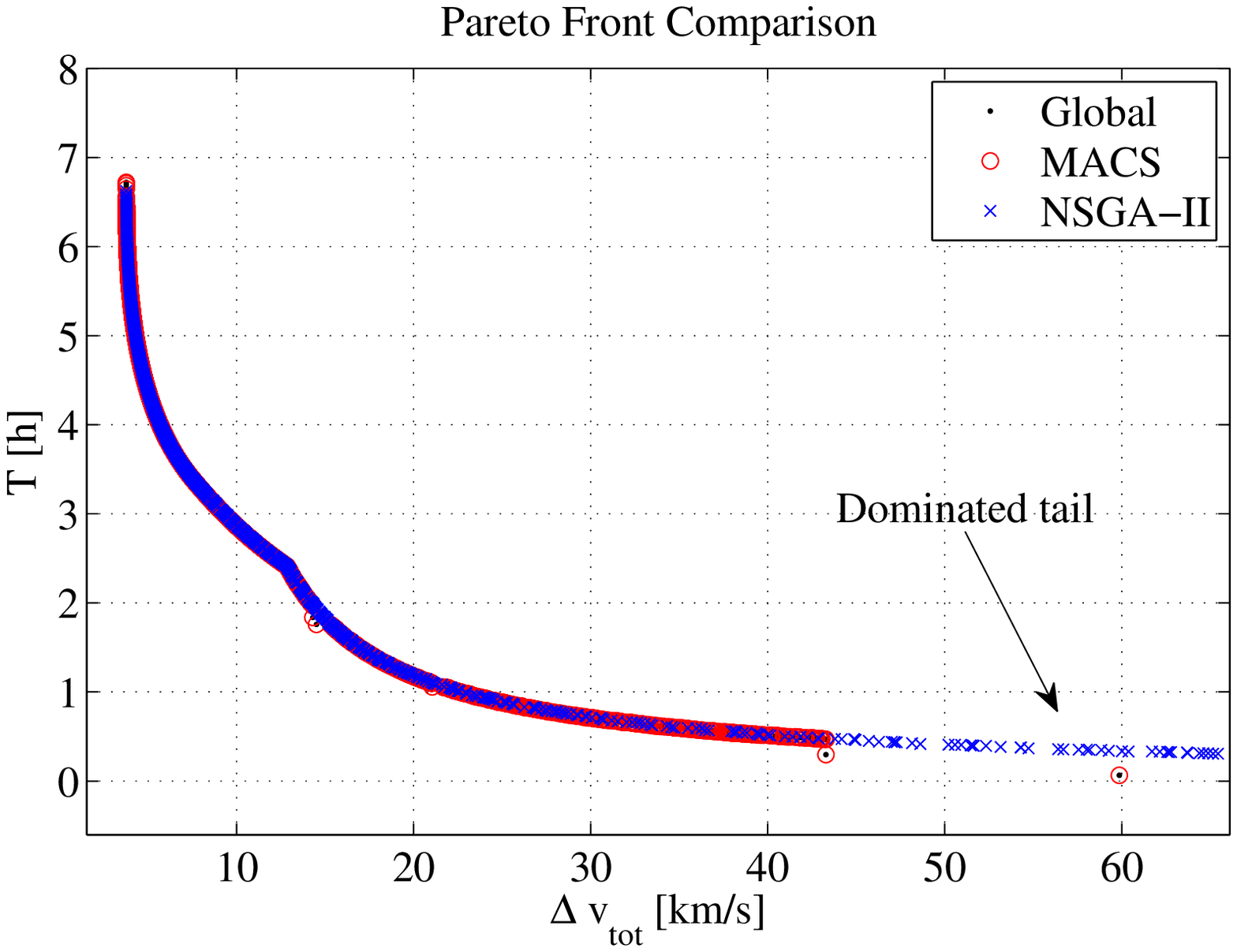}}
 \caption{Three-impulse test case: a) Complete Pareto front, b) close-up of the Pareto fronts}
\end{center}
\end{figure}

\begin{figure}[p]
\begin{center}
\subfigure[\label{fig:Cassini_front}]{\includegraphics[width=7.5cm]{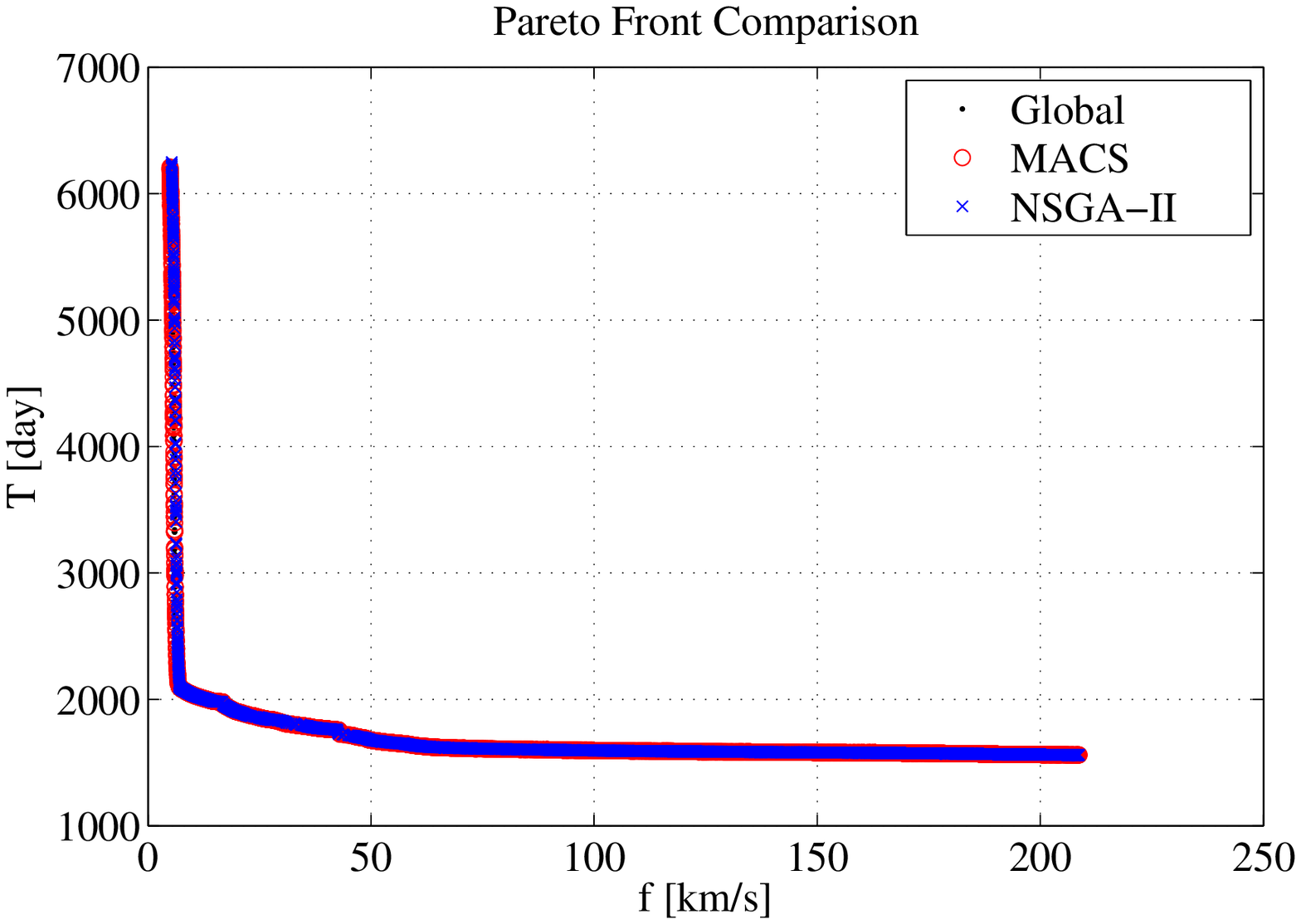}}
\subfigure[\label{fig:Cassini_front_closeup}]{\includegraphics[width=7.5cm]{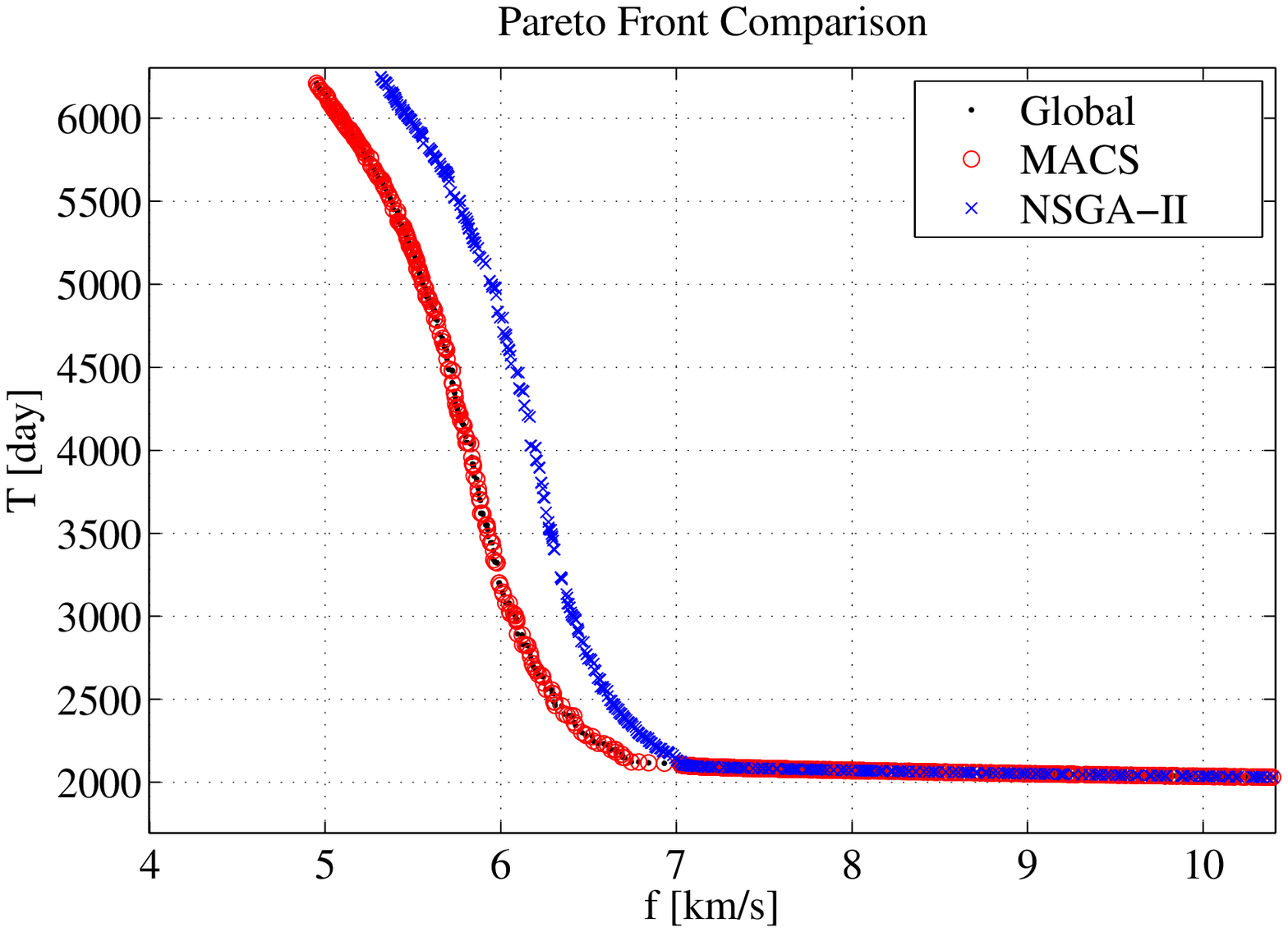}}
 \caption{Cassini test case: a) Complete Pareto front, b) close-up of the Pareto fronts}
\end{center}
\end{figure}

\end{document}